\documentclass[twocolumn]{aastex631}

\usepackage{savesym}
\savesymbol{tablenum}
\usepackage{siunitx}
\restoresymbol{SIX}{tablenum}

\usepackage{wasysym}

\DeclareSIUnit\au{AU}
\DeclareSIUnit\msun{M_\odot}
\DeclareSIUnit\mearth{M_\oplus}
\DeclareSIUnit\mmoon{M_{\leftmoon}}
\DeclareSIUnit\lem{L_\mathrm{EM}}
\DeclareSIUnit\year{yr}

\defcitealias{2019ApJEmsenhuberA}{Paper I}
\defcitealias{2021PSJEmsenhuber}{Paper II}

\def\paperone{\citetalias{2019ApJEmsenhuberA}}
\def\papertwo{\citetalias{2021PSJEmsenhuber}}

\received{2021 April 20}
\revised{2021 July 29}
\accepted{2021 July 30}
\published{2021 September 23}
\submitjournal{PSJ}

\shorttitle{Collision Chains among the Terrestrial Planets. III.}
\shortauthors{Asphaug et al.}

\begin{document}

\title{Collision Chains among the Terrestrial Planets. III. Formation of the Moon}

\correspondingauthor{Erik Asphaug}
\email{asphaug@lpl.arizona.edu}

\author[0000-0003-1002-2038]{Erik Asphaug}
\affiliation{Lunar and Planetary Laboratory, University of Arizona, 1629 E. University Blvd., Tucson, AZ 85721, USA}

\author[0000-0002-8811-1914]{Alexandre Emsenhuber}
\affiliation{Lunar and Planetary Laboratory, University of Arizona, 1629 E. University Blvd., Tucson, AZ 85721, USA}
\affiliation{Space Research and Planetary Science, University of Bern, Gesellschaftsstrasse 6, 3012 Bern, Switzerland}
\affiliation{Universitäts-Sternwarte München, Ludwig-Maximilians-Universität München, Scheinerstraße 1, 81679 München, Germany}

\author[0000-0001-6294-4523]{Saverio Cambioni}
\affiliation{Lunar and Planetary Laboratory, University of Arizona, 1629 E. University Blvd., Tucson, AZ 85721, USA}

\author[0000-0002-9767-4153]{Travis S. J. Gabriel}
\affiliation{School of Earth and Space Exploration, Arizona State University, 781 E. Terrace Mall, Tempe, AZ 85287, USA}

\author[0000-0001-6294-4523]{Stephen R. Schwartz}
\affiliation{Lunar and Planetary Laboratory, University of Arizona, 1629 E. University Blvd., Tucson, AZ 85721, USA}

\begin{abstract}
In the canonical model of Moon formation, a Mars-sized protoplanet ``Theia'' collides with proto-Earth at close to their mutual escape velocity $v_{\rm esc}$ and a common impact angle \ang{\sim45}. The ``graze-and-merge'' collision strands a fraction of Theia's mantle into orbit, while Earth accretes most of Theia and its momentum. Simulations show that this produces a hot, high angular momentum, silicate-dominated protolunar system, in substantial agreement with lunar geology, geochemistry, and dynamics. However, a Moon that derives mostly from Theia's mantle, as angular momentum dictates, is challenged by the fact that O, Ti, Cr, radiogenic W, and other elements are indistinguishable in Earth and lunar rocks. Moreover, the model requires an improbably low initial velocity. Here we develop a scenario for Moon formation that begins with a somewhat faster collision, when proto-Theia impacts proto-Earth at $\sim 1.2 v_{\rm esc}$, also around \ang{\sim45}. Instead of merging, the bodies come into violent contact for a half-hour and their major components escape, a ``hit-and-run collision.'' \textit{N}-body evolutions show that the ``runner'' often returns \num{\sim0.1}--\SI{1}{\mega\year} later for a second giant impact, closer to $v_{\rm esc}$; this produces a postimpact disk of $\sim2-3$ lunar masses in smoothed particle hydrodynamics simulations, with angular momentum comparable to canonical scenarios. The disk ends up substantially inclined, in most cases, because the terminal collision is randomly oriented to the first. Proto-Earth contributions to the silicate disk are enhanced by the compounded mixing and greater energy of a collision chain.
\end{abstract}


\section{Introduction}
\label{sec:intro}

This is the third paper in our study of nonaccretionary giant impacts and their consequences, now considering the origin of the Moon.
We begin with a synopsis of the prevailing theories for lunar formation and the underlying physics and chemistry, emphasizing the challenges and ongoing developments that motivate our effort.

\subsection{Parameters of the giant impact}

Analysis of the first lunar samples and geophysical and remote sensing data from Apollo missions \citep{2019BookCummings} led to independent arguments for the giant impact origin of the Moon \citep[][and other reviews]{2014AREPSAsphaug,2014RSPTAHartmann,2014RSPTAMelosh,2021BookCanup}.
\citet{1975IcarusHartmannDavis} proposed that instead of growing through the accumulation of planetesimals, Earth finished its accretion with several calamitous mergers involving similar-sized bodies.
Consistent with this theory, \citet{1976LPICameronWard} showed that if a Mars-sized protoplanet (eventually called ``Theia''; \citet{2000EPSLHalliday}) was accreted by proto-Earth at a typical impact angle of $\theta_{\rm coll}\sim\ang{45}$, with a velocity at contact $v_{\rm coll}\approx v_{\rm esc}$ (Equation~\ref{eq:vesc}), this would account for the high angular momentum of the Earth-Moon system.
It was further recognized that the Moon, if born at the interface of colliding terrestrial mantles, would end up with a predominantly silicate composition, explaining its small core \citep{1976LPSCGoldstein}, at most a few percent of a lunar mass $\si{\mmoon}$. This makes it a unique body in the solar system, although possibly akin \citep{2013IcarusAsphaug} to the Saturnian ice moons Tethys and Iapetus that do not have cores.

A collision between two planetary bodies of masses $m_\mathrm{tar}$ and $m_\mathrm{imp}$ and radii $r_\mathrm{tar}$ and $r_\mathrm{imp}$ makes a hyperbolic trajectory. The velocity at contact $v_\mathrm{coll}$ is thus at least the mutual escape velocity
\begin{equation}
v_\mathrm{esc} = \sqrt{2G(m_\mathrm{tar}+m_\mathrm{imp})/(r_\mathrm{tar}+r_\mathrm{imp})}
\label{eq:vesc}
\end{equation}
where $G$ is the gravitational constant.
The parameters of the canonical model are nominally $m_\mathrm{tar}\simeq\SI{0.9}{\mearth}$ and $m_\mathrm{imp}\simeq\SI{0.1}{\mearth}$ ($\gamma=m_\mathrm{imp}/m_\mathrm{tar}=1/9$) in which case $v_\mathrm{coll}\gtrsim$\SI{9}{\kilo\meter\per\second}, around twice the sound speed in geologic materials.

Shock waves from the colliding interfaces span the globes \citep{1993JGRTonksMelosh}, and other violent processes are set in motion. The energetics are especially complex for giant impacts that end in merger \citep{2020JGRECarter}, and depend on scale \citep{2010ChEGAsphaug} and on the starting composition and temperature \citep{2012ApJGenda}. The kinetic energy is the change in gravitational potential up to contact, plus the energy of original velocities and rotations. For accretions there is additionally the change in gravitational potential that plays out over hours or even days as the cores and mantles merge and attain equilibrium.
Collisional energy is applied to momentum transfer and spin-up of the target and is dissipated to internal energy by shocks, viscous heating, and friction. Heat is also consumed and released in phase transformations, as well as to surface energies in the case of expanding plumes \citep{2011EPSLAsphaug}.

Then, there are advective losses, insofar as planetary collisions are not perfect mergers. The hottest, highest-velocity debris escapes, even in slow accretions like the canonical model, removing energy from the largest remnant. High angular momentum material escapes, as well as interfacial material lost to jetting and spallation \citep[e.g.,][]{2000AREPSPierazzoMelosh}. Counterintuitively this can mean that a faster, more lossy collision can contribute \textit{less} total energy to the largest remnant of a giant impact, i.e., Earth, than a slower, more accretionary collision.

Change in enthalpy $\mathrm{d}H$ occurs in materials that start out at depth inside colliding bodies. Volumes of materials that were stable at kilobars to megabars of pressure (hydrostatic conditions in the midmantle of Theia) find themselves, hours later, inside the droplets, clumps, and filaments of the protolunar disk and the escaping debris, at a greatly reduced pressure.
Decompression makes available an enthalpy $\mathrm{d}H = V\mathrm{d}P+P\mathrm{d}V$, where $P$ is the pressure and $\rho\equiv1/V$ is the density, analogous to an ascending volcanic plume but happening globally \citep{2006NatureAsphaug,2011EPSLAsphaug}.

In summary, while the specifics are not well understood and depend on currently unknown initial states and conditions, the thermodynamic aspects of giant impacts readily account for the igneous geology of the Moon, the compelling evidence for a lunar magma ocean \citep[LMO;][]{1970ScienceWood,1985AREPSWarren}, and the depletion and fractionation of volatiles \citep{1980GCAWolfAnders}.

As for the dynamical scenario, that of a Mars-sized body straying in orbit to collide with proto-Earth, the traditional scientific context is the ``late stage'' of terrestrial planetary growth \citep[e.g.,][]{1998IcarusChambers,1999IcarusAgnor,2004IcarusRaymond}, when dozens of planet-sized ``oligarchs'' \citep{2002ApJKokubo} became gravitationally excited, collided, and \textit{overall} merged until there were four terrestrial planets and the Moon.
Three-dimensional \textit{N}-body simulations of the late stage \citep[e.g.,][]{1998IcarusChambers,2006AJKenyonBromley,2010ApJKokubo}, starting with dozens of oligarchs orbiting the Sun, can produce final architectures resembling the inner solar system, a handful of finished planets between \num{0.5} and \SI{2}{\au}, the consequence of dozens of giant impacts in an epoch lasting $\sim10$--\SI{100}{\mega\year}. These simulations are evocative but seldom end up with terrestrial planetary systems closely resembling our own. And as we describe below, the physics of collisions implemented in these models is incomplete \citep{2020ApJEmsenhuberA}.

An alternative theory \citep{2011ApJYoudinB,2015AABitschB,2016ApJChambers,2017AAMatsumura} is ``pebble accretion,'' where planets form directly from the sweep-up of small particles concentrated by aerodynamic sorting in the nebula. Pebble accretion has become the accepted framework for giant planet core formation \citep[e.g.,][]{2012AALambrechtsJohansen}; today the debate is on the extent that it applies to terrestrial planet formation, given that the process must happen in the presence of the solar nebula, in the first $\sim2$~Myr. \citet{2021SciAdvJohansen} propose, based on simulations, that Theia accreted directly as a fifth terrestrial planet, rapidly attaining a final mass of \SI{\sim0.4}{\mearth}, forming between proto-Earth (\SI{\sim0.6}{\mearth}) and Mars. In contrast to there being a late stage involving dozens of collisions, Moon formation would be the \textit{only} giant impact event, after proto-Earth and Theia somehow got dislodged from their starting orbits.

Ages of Moon formation obtained from geochemistry are broadly consistent with estimates for the duration of the late stage based on \textit{N}-body simulations and gas-free dynamics \citep{1969BookSafronov, 1980ARAAWetherill}.
The oldest geochemical age is \num{\sim10}--\SI{30}{\mega\year} after solar system formation, derived from Hf-W isotopic analysis \citep{2005AREPSJacobsen}. This may measure the core-mantle differentiation of precursor bodies; on the other hand, formation ages $\gtrsim\SI{60}{\mega\year}$ are obtained when the cosmogenic production of $^{182}$W is accounted for in rocks from the lunar surface \citep{2007NatureTouboul} .
Pristine zircons in Apollo 14 samples crystallized \SI{\sim67}{\mega\year} after solar system formation \citep{2017SciAdvBarboni}; if they solidified directly from the LMO this would place the giant impact somewhat earlier.
This is within the error bars of \SI{\sim90\pm30}{\mega\year} estimated by \citet{2014NatureJacobson} based on the abundance of accreted siderophiles after lunar solidification and the $\gtrsim\SI{70}{\mega\year}$ age \citep{2017EPSLKruijerKleine} estimated from the absence of radiogenic W variations in lunar samples.
Still later solidification ages, \num{\sim150}--\SI{200}{\mega\year} \citep{2014RSPTACarlson} or later \citep{2020SciAdvMaurice}, are obtained from combined Sm-Nd isotopic analyses of suites of crustal rocks. These may be consistent with the \SI{\sim150}{\mega\year} age obtained from U-Pb isotope systematics assuming that Pb was fractionated by the giant impact \citep{2016EPSLConnellyBizzarro}.

These estimates and others \citep[e.g.,][]{2009NatGeoNemchin} date different aspects of the Moon-forming giant impact and its magmatic aftermath. A molten Moon-sized silicate body radiating at the liquidus will solidify in \SI{<1}{\mega\year}, except that it forms an insulating crust that regulates deeper cooling \citep{1977PEPISolomon}, leading to an estimated \num{\sim10}--\SI{100}{\mega\year} solidification timescale, absent other factors. If solidification is relatively rapid \citep[e.g.,][]{2011EPSLElkinsTantonA, 2018JGREPerera} then geochemical closure might be close to the time of the giant impact.
Otherwise, LMO solidification might have been delayed by radioactive elements (incompatibles) concentrated in the residual melt or KREEP and by tidal friction during the Moon's orbital evolution, starting from a few radii away. Early, powerful tides may have sustained a partially melted state inside the LMO for $\gtrsim\SI{100}{\mega\year}$ according to \citet{2010IcarusMeyer}, which might be consistent with the theory of forced convection sustaining a lunar dynamo \citep{2011NatureDwyer}.

In any case, we must be cautious in tying LMO solidification and the associated closure ages, directly to the Moon's formation. A Ceres-sized asteroid colliding with the Moon after it had solidified (even a late-returning remnant of the giant impact itself as discussed below) would create a gigantic basin whose melt volume would exceed the crater volume \citep{1989BookMelosh}, producing a new magma ocean spanning the impacted hemisphere. Crystallization sequences of lunar samples might in that case date a later, regional solidification \citep{2011NatureBorg} and not the Moon-forming giant impact.

\subsection{Collisional dynamics}

A giant impact starts with a dynamical perturbation that leads to a collision and its aftermath. The collision begins hours ahead of contact, when the planets approach within their tidal influence, several planetary radii apart. The smaller body is subjected to increasing tidal deformation and torque, which changes its cross section and rotation, and hence changes the impact parameters---considerations that are astrophysical as much as geophysical.
It is therefore natural that the modern understanding of the giant impact process began with the application of 3D stellar astrophysics simulations, modified for geophysical equations of state \citep[EOSs;][]{1986IcarusBenz} --- the smoothed particle hydrodynamics (SPH) method from which our present code derives \citep[see Section~\ref{sec:methods} and][hereafter \papertwo{}]{2021PSJEmsenhuber}.

Early suites of SPH modeling experiments \citep[e.g.,][]{1991IcarusCameronBenz,1997IcarusCameron} showed that massive silicate disk formation is favored by ``graze-and-merge'' collisions \citep{2010ApJLeinhardt}; these are common for giant impacts within a few percent of the escape velocity, at impact angles in the range of $\theta_{\rm coll}\gtrsim\ang{30}$. These off-axis accretions bring significant angular momentum and, for most angles, merge the metallic cores so that the disk is made almost entirely of silicates.
Calculations using the same code at higher resolution \citep{2001NatureCanup} confirmed that the scenario postulated by \citet{1976LPICameronWard}, a collision by a Mars-sized Theia at \ang{\sim45} at close to $v_{\rm esc}$ into an almost-finished proto-Earth, led to outcomes that could match the mass distribution, composition, and angular momentum of the Earth-Moon system.

The projectile is not stopped abruptly in a giant impact the way a bullet is stopped in a block of wood, or an asteroid striking the Earth. Giant impacts are similar-sized collisions, meaning they are usually significantly off-center, like a rugby ball hitting a basketball, or two pool balls. A direct hit is not typical. When the sine of the impact angle exceeds the fractional radius of the metallic core, i.e. $\sin^{-1}(1/2)=\ang{30}$ in the case of chondritic compositions, the colliding cores do not touch along the projected impact vector. This is more head-on than the average impact angle of \ang{45}, so that most giant impacts are grazing \citep{2010ChEGAsphaug} by this definition.

The impact angle specified in the canonical model \citep{1976LPICameronWard}, about \ang{45}, is required to provide sufficient Earth-Moon angular momentum, assuming zero rotation to start with. As noted, most of the projectile ``misses'' the target \citep[e.g.,][]{2016IcarusMovshovitz}, and the bodies do not slow down enough to end in immediate accretion. What happens next depends on the impact velocity $v_{\rm coll}$. If close to $v_{\rm esc}$, most giant impacts are graze-and-merge collisions like the canonical case, drawn-out pinwheel mergers that are good at stranding silicates in orbit. If somewhat faster, the projectile can escape, continuing as a ``runner'' discussed below.
The latter ``hit-and-run'' collisions are generally poor candidates for Moon formation, because they fail to accrete the angular momentum, although \citet{2012IcarusReufer} identified some potential scenarios.

Accretion at higher $v_{\rm imp}$ can occur, but this requires a relatively head-on impact to decelerate the projectile so as to end in capture, which in turn brings insufficient angular momentum to explain the Earth-Moon system. As for lower velocities, the database of satellite-forming giant impacts includes examples of direct capture, when just the right fraction of angular momentum is lost and dissipated. In this manner a slow, highly grazing projectile can end up relatively intact and in orbit. But this also captures the core into orbit, so is contrary to the composition of the Moon but is consistent with Pluto-Charon formation \citep{2011AJCanup}.

For the canonical giant impact parameters the runner does not escape but loops back for a reimpact several hours later, along with shredded outer-mantle and crustal/oceanic debris. The re-impact is, in turn, a slower grazing merger, although sometimes it can can fly over the target in the loop-back orbit, which can then lead to potential direct-capture scenarios. More often, the gravitationally bound cores of the colliding bodies rapidly combine into a lopsided, spinning central body that transfers gravitational torques to protodisk silicates, and to protolunar clumps, creating one or more tidal spiral arms that expand upon release from shock and from hydrostatic pressure inside Theia.

In summary, giant impact accretions are complex processes whose detailed outcomes depend sensitively on the mass ratio, impact angle, velocity, composition, and thermal state of the colliding bodies. Yet despite this complexity, numerical simulations of graze-and-merge collisions starting with canonical giant impact parameters often lead to reproducible outcomes that are consistent with Moon formation, although the specific results depend on the numerical method and resolution as discussed below.

The canonical scenario is widely appealing. It is compatible with astrophysical predictions of oligarchic growth \citep[e.g.,][]{2002ApJKokubo}, where two planetary embryos get into crossing orbits owing to late gravitational perturbations and collide. It satisfies the constraints of Earth-Moon dynamics, producing a stable disk that is more massive than the Moon and a disk and an Earth-Moon system with the right angular momentum \citep{2001IcarusCanup}.
The disk is made of molten silicates, satisfying the riddle of the Moon's small iron core \citep{1976LPSCGoldstein}, and providing a thermodynamic explanation for the LMO.
A hot start might also explain the volatile-poor nature of the Moon, although volatile escape is perhaps not as significant as originally thought \citep[see][]{2018EPSLNakajimaStevenson,2018JGRELock}, and the Moon is not as volatile depleted as originally thought \citep[e.g.,][]{2011ScienceHauri}.

Giant impacts play a significant role in planetary habitability \citep[e.g.,][]{2007SSRvZahnle}, so it is important to identify the scenario that made the Moon. Major outcomes (such as hit-and-run or graze-and-merge) are sensitive to the impact angle and velocity, and these parameters are stochastic, so overall there is a predicted diversity of giant impact outcomes that could account for the diversity of potentially habitable planets \citep{2010ChEGAsphaug}.
If giant impacts were rare, or if Moon formation required special conditions, this might explain why Venus lacks a moon and rotates slowly, and lacks a magnetic field, and is overall ``un-Earth-like.'' One can speculate that perhaps the final giant impact into Venus was retrograde, robbing angular momentum and destroying an existing moon. Or, Venus never suffered a giant impact \citep{2017EPSLJacobson,2021SciAdvJohansen}, or it accreted from systematically different giant impacts than the Earth (the subject of \papertwo{}). Giant impact modeling does not answer these questions but allows us to better understand and evaluate the assumptions and quantitative implications of proposed scenarios.

\subsection{Limits of simulations}

Meaningful simulations of giant impacts must accurately model multiple stages of physics: the dynamics leading up to the collision, the formation and release of impact shocks and compressions, the global long-range interactions of self-gravity, and melt-vapor thermodynamics which depends sensitively on nonlinear EOSs and precise treatment of energy and entropy by the code. The EOS, in turn, assumes that we know the compositions and constitutive properties of the colliding bodies.

As for modeling more complicated planets, such as an ocean on the proto-Earth, this is resolvable in 1D \citep{2005NatureGendaAbe} and 2D, but not yet in routine 3D simulations of giant impacts. An ocean would have to be more than a few hundred kilometers deep to be adequately resolved in our current models. A target with such a large water fraction would respond quite differently to a giant impact for reasons of moment of inertia and EOS \citep[e.g.,][]{2018ApJHaghighipour}. For now we stick to the well-studied two-component planets (rock and metal), focusing on the physics in 3D, and relying on thermodynamically consistent EOSs for forsterite and iron.

SPH is accurate in computing global self-gravity, and unlike most grid-based codes, it conserves angular momentum precisely. This, as well as its relative ease of use and analysis, has made it a standard method for simulating giant impacts. Modern implementations of SPH are accurate in energy conservation to better than \SI{1}{\percent}. As with any hydrocode, the results are limited by resolution, since self-gravity and off-axis dynamics require 3D evolution for several gravitational timescales, up to days. Artifacts such as numerical viscosity and numerical tension can masquerade as physical, and sparse-particle regions can experience fictional shear forces. Heating \cite[e.g.,][]{2021ApJLGabrielAllen-Sutter} and disk stability \cite[e.g.,][]{2016ApJRaskinOwen} are sensitive to the artificial viscosity used to resolve shocks.

It is encouraging that simulations of the canonical model by various groups \citep[e.g.,][]{2012IcarusReufer,2015EPSLNakajimaStevenson}, even using diverse techniques and resolutions \citep{2009ApJMarcus,2013IcarusCanup,2016JGREBarr}, often give comparable results for the mass, composition, and angular momentum of the protolunar disk, for similar initial parameters.
However, this is not universally the case, and disagreements might influence the viability of the hypothesis. For example, \citet{2012JGREKraus} argue for significantly greater vapor production compared to the ``flying magma ocean'' predicted earlier \citep{1987AREPSStevenson}, factors that also depend on the thermal state of the target \citep{2012ApJGenda, 2019NatGeoHosono}. Gravitational instability (coagulation) and the onset of shocks that might disperse accreting regions of the disk both depend on the local sound speed in the disk. The phase, pressure, and temperature of the postimpact disk may thus determine whether the Moon can accrete before the material disperses. Multi-million-zone Eulerian simulations by \citet{2006ApJWada} end with a disk comparable to other simulations but indicate that it would be subject to internal shocks that could severely frustrate the growth of the Moon.

\begin{figure*}
	\centering
	\includegraphics{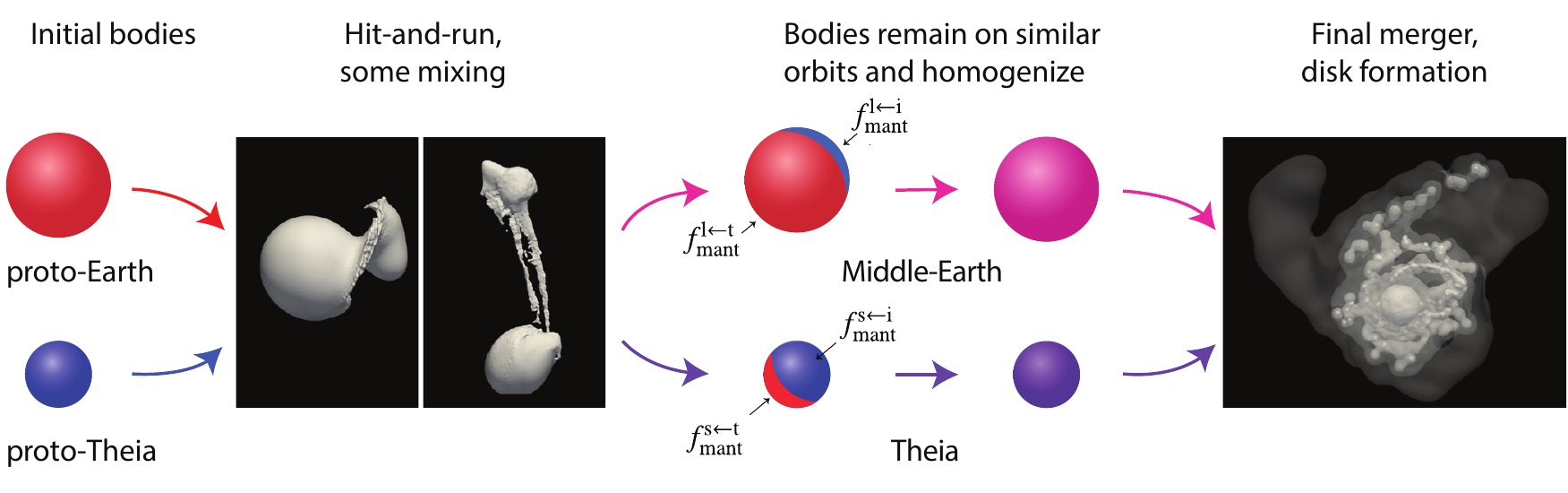}
	\caption{Schematic representation of the hit-and-run return scenario for Moon formation. Grayscale images are renderings of representative SPH results of the two giant impacts. The first is a hit-and-run, producing largest ($l$, aka middle-Earth) and second-largest ($s$, the runner, aka Theia) remnants. The second collision is slower, a graze-and-merge that forms a massive disk. Colors indicate mixing, where $t$ and $i$ refer to target and impactor (proto-Earth and proto-Theia), respectively, per Eq.~\ref{eq:dft2}.}
	\label{fig:schema}
\end{figure*}

A comparative study by \citet{2013IcarusCanup} using different hydrocodes (CTH and SPH) at various resolutions (e.g., \num{e4} to \num{e6} particles) found first-order similarities in terms of disk mass and angular momentum, but they identified substantial structural differences. Some simulations ended with one or two massive protolunar clumps totaling about one lunar mass, embedded in a less massive disk, while others ended with a clump-free disk. They found no trend in clumping with numerical resolution. According to \citet{2017PASJHosono} numerical predictions for circumterrestrial postimpact structures have not converged even with millions of particles.

Even when giant impacts are properly and accurately computed by a code, they are sensitive to relatively minor changes in impact angle, velocity, composition, and mass ratio \citep[roughly in that order; see][]{2020CACTimpe}. Strong variation is observed within common ranges of those parameters \citep{2020ApJGabriel}.
Predictive sensitivity extends to the static compressibility of the EOS \citep{2020AAWissingHobbs}. ANEOS for forsterite and iron used in our ongoing research (see Section~\ref{sec:methods}) gives a more realistic and complete treatment of shocked material than the simple Tillotson EOS that was used in earlier studies \citep[e.g.,][]{2001NatureCanup,2004ApJAgnor}.
Yet despite its shortcomings \citep{2019AIPStewart}, an Earth-mass planet constructed using Tillotson can end up with approximately the correct target radius and central density, whereas planets constructed using ANEOS end up a fraction too compressed. This makes a giant impact slightly more grazing ($\sin\theta_{\rm coll}=b/r_{\rm tar}$) for a given impact parameter $b$, and the bodies start and end more gravitationally bound.
It therefore remains quite challenging to relate differences in giant impact outcomes to differences in material physics, versus differences in implementation.

\subsection{Isotopic contradictions}

The canonical model is a slow ($v_{\rm imp}\approx v_{\rm esc}$), near-perfect merger with an accretion efficiency $\xi\gtrsim0.98$ in most simulations, where
\begin{equation}
\xi=(m_\mathrm{lr}-m_\mathrm{tar})/m_\mathrm{imp}
\label{eq:xi}
\end{equation}
equals 1 for perfect merger. Here $m_\mathrm{lr}$ is the mass of the final largest remnant (Earth in this case) and $m_\mathrm{tar}>m_\mathrm{imp}$ are the target and projectile masses (i.e. proto-Earth and Theia).
The Earth ends up being made out of both bodies proportional to their contributions, so about $1/(1+\xi\gamma)$ proto-Earth.
The disk, though, ends up deriving mostly from Theia, specifically the mantle fraction that is approximately opposite its initial contact with the Earth. This distal mid- to outer mantle of Theia has the greatest captured angular momentum and so contributes more than two-thirds of the protolunar disk in simulations \citep[e.g.,][]{2001NatureCanup,2012IcarusReufer}.
The fraction $1-\xi$, also mostly from Theia, remains in heliocentric orbit, unaccreted by Earth for now.

The provenance of protolunar disk material is a serious problem for the canonical model. The Moon was not blasted out of Earth's mantle, as is often envisioned; it is instead a lossy collisional capture of a silicate portion of Theia.
Lunar rocks should therefore be readily distinguishable from Earth rocks, in the way that meteorites from asteroids or from Mars are distinct from each other and from Earth in isotopic composition. Theia would be distinct from proto-Earth according to oligarchic growth \citep{2002ApJKokubo}, where embryos are born in well-separated feeding zones of the nebula, where temperature, pressure, and composition variations would give rise to isotopic differences \citep[e.g.,][]{2012EPSLBurkhardt}.

Yet Apollo samples from the nearside, as well as lunar meteorites from all over the Moon, have oxygen isotopic ratios ($\delta^{17}O$) that are indistinguishable from Earth \citep{2001ScienceWiechert,2007EPSLSpicuzza,2010GeCoAHallis}, even to a few ppm \citep{2016ScienceYoung}. Other rock-forming species are indiscernible as well \citep{2014PTRSADauphas, 2017NatureDauphas}, for instance, Ti \citep{2012NatGeoZhang} and Cr \citep{2010GCAQin}, which have quite different thermophysical and petrological behaviors and associations; radiogenic W \citep{2007NatureTouboul}; and Si \citep{2012GCAArmytage}.
It presents a striking dilemma: if the canonical model is correct, how could Theia have been so alike the Earth?

The most convenient explanation would be that Theia and proto-Earth were born in the same feeding zone, as this would also be consistent with a low-velocity collision.
The idea that Theia was a Trojan planet that co-accreted in a 1:1 resonance with proto-Earth \citep{2005AJBelbrunoGott} has been ruled out \citep{2016IcarusKortenkampHartmann}, but the idea of Theia being a dynamical neighbor of proto-Earth has some basis \citep{2015IcarusQuarlesLissauer,2015NatureMB}. Other arguments are to the contrary \citep{2015IcarusKaibCowan}.
The explanation would require Theia to be born so close in time and space to proto-Earth as to have their compositions indistinguishable, yet to have their collision deferred by $\sim\SI{100}{\mega\year}$ after their formation. Perhaps the inner Solar System is more isotopically homogeneous than meteorites suggest; until we have samples from Mercury and Venus this dilemma will be mired in speculation.

The same-feeding-zone explanation is not without geochemical paradoxes of its own. If the feeding zones were the same, then Si would have started off the same. But deep inside proto-Earth, Si would have partitioned more effectively to the Fe-Ni core under pressures that were 10 times greater than inside Theia. According to \citet{2012GCAArmytage}, there would be heavier Si remaining in Theia's mantle and hence the Moon, yet lunar Si is the same.
Another puzzle is FeO, which appears to be significantly more abundant in lunar mantle than in Earth. This can be accounted for by FeO disproportionation at higher pressure \citep{2008AREPSFrostMcCammon}, and possibly by changes in oxidation in the aftermath of giant impacts \citep{2021PSJCambioni}. It could also indicate a more oxidized Theia \citep{2019NatAsBudde}; however, in that case an identical isotopic reservoir is even less likely. And lastly, there is radiogenic W, where non-cosmogenic $^{182}$W/$^{184}$W ratios are the same in rocks from the Earth and the Moon \citep{2007NatureTouboul} and could require coincident differentiation in Theia and proto-Earth.

The sampled lunar crust is not the bulk Moon \citep{2014EPSLGross}, and moderately volatile elements such as Rb and Zn show evidence for isotopic fractionation in crustal rocks \citep{2019ApJNieDauphas}.
Contrary to prior studies, \citet{2020NatGeoCano} report that lunar rocks do have small variations in $\delta^{17}O$ around an average value that matches that of Earth. They argue that their reported variations correlate systematically with petrology, and they propose that the interior of the Moon has heavier oxygen than the Earth (being mostly Theia), while the crustal suite happens to overlap the bulk silicate Earth owing to fractionation between the magma ocean and an escaping lunar silicate atmosphere.
In any case, isotopic composition is perhaps not so straightforward a constraint on scenarios of Moon formation.

\subsection{Mixing, Diffusion, and Layering}

One way to reconcile the Earth-Moon isotopic similarity is to invoke widespread compositional diffusion between the protolunar disk and the postimpact Earth's silicate atmosphere and magma ocean. This would apply to any mechanism of Moon formation, not just the canonical model, and would be more effective in alternative models that involve greater collisional heating and mixing.

\citet{2007EPSLPahlevan} calculate that proto-Earth and a vapor-rich protolunar torus could have approached \SI{90}{\percent} diffusive equilibrium in $\sim$100~yr. This is longer than the $\sim$0.1-1~yr accretion timescale obtained by \citet{1997NatureIda} and \citet{2000IcarusKokuboB}, although in those approaches the Moon forms from an \textit{N}-body disk of large solid particles subject to gas-free coagulation. Even for the relatively gentle canonical model, simulations predict \SI{>20}{wt\percent} silicate vapor production \citep[e.g.,][]{2014IcarusNakajimaStevenson}. A melt-vapor torus would coagulate on longer timescales, limited by cooling \citep{1987AREPSStevenson}.

Diffusion in this manner requires a material connection between the disk and postimpact Earth.
Simulations of the canonical model produce a postimpact disk that straddles the Roche limit $R_{\rm Roche}$, about 2.9 Earth radii for silicate moonlets, and that spreads onto the Earth. Bodies coagulating interior to $R_{\rm Roche}$ tend to shear apart; this could lead to heating, fluidization, and viscous spreading in the model of \citet{2012ApJSalmon}.
Exterior to $R_{\rm Roche}$, moonlets would grow under stable conditions and rapidly accrete to form a massive satellite \citep[e.g.,][]{2000IcarusKokuboB}.

Oxygen represents almost \SI{40}{\percent} of the mass of silicates, so its widespread diffusion would ultimately require the exchange of most of the angular momentum in the disk. Angular momentum transfer would lead to the dispersal of the disk \citep{2014RSPTAMelosh}, a contradiction. Furthermore, wholesale oxygen exchange between Earth's early mantle and the disk requires a subsequent explanation for why the Moon ended up several-fold depleted in water \citep{2011ScienceHauri}. One possibility is the subsequent escape of volatiles at the Hill radius of the accreted Moon \citep{2021IcarusCharnoz}.

A compromise advocated by \citet{2012ApJSalmon} is that the massive inner disk would more rapidly attain isotopic equilibration with the Earth. This equilibrated disk would spread, a fraction condensing into bodies that would be accreted by the proto-Moon outside $R_{\rm Roche}$. The deep mantle and core of the Moon would be made mostly of Theia, while the outer mantle and crust (that we sample) would be made of more Earth-like inner-disk material. This would contrast with how lunar formation geology is usually interpreted, a complex plagioclase cumulate above a global magma ocean \citep[e.g.,][]{1985AREPSWarren}.

Another idea for equilibration invokes a series of not-quite-giant impacts that each launched mostly proto-Earth-derived silicates into orbit \citep{2017NatGeoRufu}. If impacting at high enough velocity and working together, these could build up the Moon sequentially.
This approach has the benefit of averaging out the heterogeneities contributed by individual projectiles, allowing for a realistically diverse bombarding population.
The critical problem is that each event would have to add to the previous disk or proto-Moon and not erode it, or cause it to de-orbit, implying that the collisions would somehow have to be aligned in angular momentum.

\subsection{System angular momentum}

An original tenet of the giant impact hypothesis \citep{1976LPICameronWard} is that the Earth-Moon system ended with approximately the same angular momentum $L_{\rm EM}$ as it has today. The Moon currently accounts for $\sim4/5$ of the system angular momentum, acquired from Earth during its outward tidal migration. Projecting back in time to Moon formation, \citet{1879RSPTDarwinB} calculated that most of $L_{\rm EM}$ was in the early Earth that spun with a period $P_{\rm rot}\approx5$ hours. This quantity is generally thought to be conserved during coupled Earth-Moon evolution.

This is different from the related question of how much angular momentum $L_{\rm disk}$ has to end up in the disk orbiting the postimpact Earth at the end of a collision. A minimum is sometimes taken to be $\sim0.18 L_{\rm EM}$, that of a \SI{1}{\mmoon} body in circular orbit at the Roche limit \citep{2004IcarusCanup}. Although this can be useful as a comparative criterion, $L_{\rm disk}$ continues to evolve beyond the time frame of SPH simulations owing to the mass asymmetry of the spinning postimpact Earth and the ongoing evolution of massive clumps.

Like isotopic abundances, the system angular momentum $L_{\rm EM}$ has been taken as a fundamental quantitative constraint on giant impact scenarios. But this approach is not straightforward either.
As the Moon migrated out, the Sun increased in relative gravitational influence. At some point the Moon encountered the evection resonance \citep{1998AJToumaWisdom}, where its orbital precession around the Earth equals \SI{1}{\year}. The angle between the Sun and the line of apsides would be constant, so a torque would accumulate that could drain the Earth-Moon angular momentum \citep{2012ScienceCuk}.
If the Moon migrated slowly, then it could have been trapped in the evection for some time. Slow migration, in turn, requires low tidal friction and thus severely constrains the Earth's interior and crustal evolution \citep{2015EPSLZahnle}.

\citet{2015IcarusWisdom} showed that capture into evection can occur only for a narrow range in relative tidal dissipation inside Earth and the Moon; otherwise, orbital eccentricity increases and the Moon escapes the resonance. \citet{2020JGREWard} obtained the result that evection could not be maintained for most tidal parameters. It is possible that other perturbations \citep{2016NatureCuk} and quasi-resonances \citep{2017IcarusTian,2020JGRERufuCanup} may have applied \citep[see, however,][]{2020PNASTianWisdom}. So with significant caveats the door remains open to higher angular momentum scenarios.

One such scenario is the merger, at close to $v_{\rm esc}$, of equal-mass semi-Earths. \citet{1997IcarusCameron} demonstrated the idea but did not consider it further, as it accretes twice the $L_{\rm EM}$. It was examined more carefully by \citet{2012ScienceCanup} in the context of isotopic similarity. Its appeal is that if the merging bodies are very close in bulk composition (e.g., both chondritic), then Earth and the Moon would accrete in equal fraction out of each body, guaranteeing isotopic similarity no matter how diverse their origins. However, if the twin planets varied by several percent in mass or moment of inertia, or rotated much differently, then the collision would be lopsided and the compositions would be distinct, so the proposed solution applies for a quite narrow range of cases. Still, equal-mass collisions are 2-3 times as energetic as canonical collisions \citep{2020EPSLLock}, so postimpact mixing, diffusion, and layering described above would be more effective at reducing isotopic differences following the collision.

In terms of evidence, there are no half-Earth planets remaining, but two Theia-mass planets, Mars and Mercury, that could be leftovers \citep{2014NatGeoAsphaug}.
But absence of evidence is not evidence of absence, especially in this case, as the semi-Earths would be beneath our feet. By whatever process two equal-mass progenitors may have come about, the idea faces two major dynamical challenges. One is the high $L_{\rm EM}$ of the merger. The other is to explain how two major planets would be perturbed into colliding orbits $\sim\SI{100}{\mega\year}$ after their formation. In the scenario of pebble accretion \citep[e.g.,][]{2021SciAdvJohansen} they would form on rather circular orbits at least \SI{1/3}{\au} apart, though there may be a tendency for resonant pairs \citep{2019AALambrechts}.

If the Earth-Moon system can shed angular momentum, then perhaps the proto-Earth could have been spinning 10 times faster than today. In this case the target's equator would be already almost escaping owing to centrifugal forces, so that Moon formation could be an ``impact-triggered fission,'' the scenario proposed by \citet{2012ScienceCuk}. In their Figure~1 a projectile impacts the equator of the oblate proto-Earth at \SI{20}{\kilo\meter\per\second}, about twice $v_{\rm esc}$, somewhat counter to its rotation ($\theta_{\rm coll}=-\ang{20},\varphi=\ang{180},P_{\rm rot}=2.3$ hours). The equator spins toward the projectile at up to \SI{\sim10}{\kilo\meter\per\second}, greatly increasing the impact energy while lowering the preimpact binding energy. The result is a global-scale explosion and expansion \citep[``synestia'',][]{2018JGRELock}, much of which recondenses onto the spinning core, leaving a vaporized torus of Earth-derived material to make the Moon.

Theia must come from the outer solar system to have such high velocity \citep{2018MNRASJackson}, and from there the delivery probabilities are low \citep[e.g.,][]{2000MAPSMorbidelli}. The projectile, twice the mass of Ganymede, would be from a lost or undiscovered population. On the other hand, the composition of Theia would be inconsequential because the projectile is dispersed; the scenario perfectly addresses the isotopic similarity. It begs the question of how proto-Earth could have spun up so fast in the first place. Terrestrial planets end up spinning relatively slowly under pebble accretion \citep{2020IcarusVisser}. As for late-stage accretion, \citet{1999IcarusAgnor} found that perfect merging would lead to fast-spinning planets; however, their conclusion was that perfect merging is a wrong assumption. To spin proto-Earth to $P_{\rm rot}<\SI{3}{\hour}$ would seem to require one or more grazing mergers ahead of the proposed collision.

\section{Hit-and-run return}

Giant impacts at $\gtrsim1.2 v_\mathrm{esc}$ are common in \textit{N}-body studies of late-stage planet formation and in classical theory \citep[e.g.,][]{1980ARAAWetherill}, yet relatively few studies have considered Moon formation in this velocity range \citep[e.g.,][]{2012IcarusReufer,2017NatGeoRufu}. These are in a different regime than canonical graze-and-merge scenarios but much slower than impact-triggered fission.

Giant impacts at higher velocity convey more angular momentum than the canonical model, as well as more kinetic energy, so in principle they are promising for Moon formation.
But for a similar-sized projectile to convey the requisite angular momentum at $\gtrsim1.2 v_\mathrm{esc}$, the impact angle must still be $\gtrsim\ang{30}$, so the target does not manage to capture the angular momentum, although there is more of it coming in.
Such hit-and-run collisions can at first resemble graze-and-merge events---the energetics and dynamics are initially quite similar. But instead of returning on a bound orbit, the impactor (now runner) escapes, often barely. Some of the projectile may be accreted in a hit-and-run, and there can be significant mixing during the half hour of violent contact. But the high angular momentum material that is needed to explain $L_{\rm EM}$ and the protolunar disk is lost.

The idea that Theia might escape from a giant impact presents another opportunity for explaining the isotopic similarity, as in the scenario by \citet{2012ScienceCuk} but at lower energy.
\citet{2012IcarusReufer} demonstrate that relatively head-on (less than \ang{\sim40}) hit-and-runs by massive projectiles can dredge up Earth-derived mantle and retain it as a protolunar disk, with most of Theia's contribution escaping. Focusing on mass ratios about twice that of the canonical model ($\gamma\sim1:5$) and a range of projectile-target compositions including icy and metallic bodies, they identify scenarios that improve by a factor of two the isotopic fit. Their scenario applies to a narrow range of parameters, however, and they do not end up with quite a lunar mass in the disk (although using the stricter periapse criterion; see below). And a hit-and-run collision leaves open the question, \textit{what happens to the runner?}

\subsection{Accretion efficiency}

Moon formation is such a persistent problem in computational physics in part because of the precision that is required to obtain a meaningful solution. A realistic calculation must resolve the dynamics and thermodynamics of the hottest, most tenuous few percent of the total colliding mass and determine its fraction that neither escapes nor is accreted but ends up in stable orbit around the spinning Earth.

A much simpler and more robust computation is the accretion efficiency $\xi\leq1$ (Equation~\ref{eq:xi}) of a giant impact, whose determination focuses on the coldest, least tenuous, most massive components of a collision outcome that are gravitationally bound. The major bound masses are the postimpact target and the runner, if there is one; these are reliably obtained to within a few percent in comparative simulations except around the sensitive transition between graze-and-merge ($\xi\approx1$) and hit-and-run ($\xi\approx0$).

Accretion efficiency discriminates strongly between scenarios of Moon formation. For example, $\xi\sim0$ for \citet[][; hit-and-run]{2012IcarusReufer}, $\xi<0$ for \citet[][; fission]{2012ScienceCuk}, and $\xi\sim1$ for the canonical model.
Perhaps because of the decades of focus on the latter, and for the simplicity of the assumption, perfect merging ($\xi=1$) has been a common expediency in \textit{N}-body studies of planet formation: two bodies come in (often with an ``expansion factor'' so they collide even if they miss by several radii) and a combined body goes out with the equivalent mass and momentum.
\citet{1999IcarusAgnor} examined the implication of this assumption, tracking the accretion of angular momentum along with the accretion of matter.
They found that planets end up spinning well beyond the disruption limit assuming perfect merging, with Earth-mass planets acquiring rotation periods shorter than 1~hr. This is impossible \citep{1969BookChandrasekhar}; the reality is that accretion is inefficient for similar-sized collisions when the product of impact velocity and impact angle is large.

For planets of a given composition, such as differentiated chondritic bodies, $\xi$ is approximately a function of the dimensionless parameters
\begin{equation}
\xi=\xi(\gamma,v_{\rm coll}/v_{\rm esc},\theta_{\rm coll})
\end{equation}
where $\gamma=m_\mathrm{imp}/m_\mathrm{tar}$ is the mass ratio and $\sin(\theta_{\rm coll})$ is equivalent to the normalized impact parameter $b/(r_{\rm imp}+r_{\rm tar})$.
Composition, as represented by a variable core mass fraction $Z$, introduces another dimensionless parameter that affects collisions systematically \citep{2020CACTimpe}.

Departures from scale invariance occur because EOSs are nonlinear, and because constitutive properties (strength, friction, porosity) are disproportionately significant at smaller masses \citep{2018IcarusEmsenhuber}.
In collisions between accreting planetesimals (``small giant impacts'') $v_{\rm esc}$ is subsonic, so the linear (e.g., elastic) response of the EOS matters most, plus crushing behavior and friction \citep{2015PSSJutzi} that depend on gravity and pressure.
The nonlinear EOS responses become dominant at larger scales for two main reasons: the escape velocity is higher, so the collisions are faster and the shocks are more intense \citep{2020JGRECarter,2021ApJLGabrielAllen-Sutter}; and more massive bodies begin and end a collision more centrally condensed, changing the gravitational energetics, making it increasingly difficult to exhume core material, and increasing the preponderance of hit-and-run collisions \citep{2020ApJGabriel}.

The mass ratio $\gamma$ is associated with each Moon-formation scenario: $\sim1/9$ in the canonical model, $\sim1/20$ in the model of \citet{2012ScienceCuk}, $\sim2/3$ for pebble accretion \citep{2021SciAdvJohansen}, and $\sim1$ for semi-earths. The closer in mass, the more physically grazing a collision for a given impact angle, making hit-and-runs more common for higher $\gamma$.
As for impact angle $\theta_{\rm coll}$, this parameter is stochastic for most giant impacts \citep[e.g.,][]{2019ApJEmsenhuberB} and follows the familiar \citet{1962BookShoemaker} probability distribution $\mathrm{d}p(\theta_{\rm coll})=\sin{2\theta_\mathrm{coll}}\mathrm{d}\theta_\mathrm{coll}$. Half of giant impacts are between \ang{30} and \ang{60}. The present study focuses on \ang{\sim45} collisions, as they have maximum probability.

Impact velocity $v_{\rm coll}=\sqrt{v^2+v_{\rm esc}^2}$, where $v$ is the relative velocity before the encounter, which depends on the starting orbits of the colliding bodies. In a self-stirred system (not excited by external perturbers) $v$ is also stochastic, with an average value that may be estimated by the statistics of encounters \citep[e.g.,][]{1969BookSafronov}. In the absence of nebular gas and particle drag, mutual gravitational encounters increase random velocities until, over time, they might approach the escape velocity of the major perturbers \citep{1980ARAAWetherill}, in which case $\langle v_{\rm coll} \rangle \rightarrow \sqrt{2} v_{\rm esc}$ depending on the distribution, with slower and faster collisions as outliers. Gravitational drag causes the relative velocities of the major bodies to decrease with an increasing mass fraction in planetesimals \citep{2006IcarusOBrien,2013IcarusChambers,2015IcarusKaibCowan}.

The collisions that characterize the late stage depend sensitively on whether $\langle v_{\rm coll} \rangle$ is $1.1 v_{\rm esc}$, $1.2 v_{\rm esc}$, or faster, as this represents the transition from accretion to hit-and-run for expected mass ratios \citep[see, e.g.,][and other studies]{2012ApJLeinhardt,2012ApJGenda}.
This transition was first explored by \citet{2004ApJAgnor} on the basis of dozens of SPH simulations using the same code and EOS as \citet{2001NatureCanup} but starting with Mars-mass progenitors. They noted a precipitous drop in $\xi$ with impact angle over the velocity range associated with moderate self-stirring, showing that realistic accretion efficiency must be accounted for in \textit{N}-body studies of planet formation.

\begin{figure*}
	\centering
	\includegraphics[width=2\columnwidth]{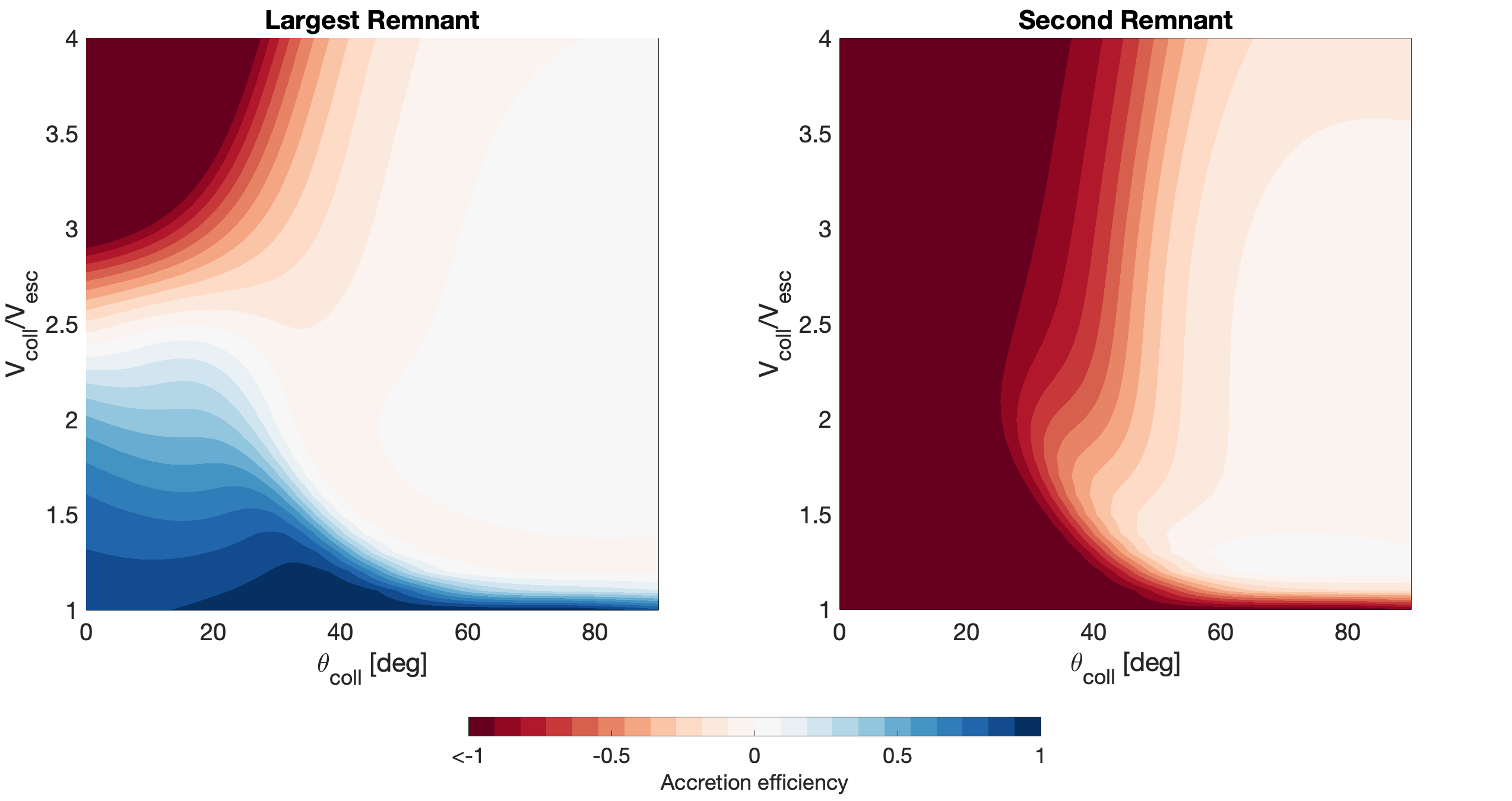}
	\caption{Left: accretion efficiency $\xi$ of giant impacts (Equation~\ref{eq:xi}) plotted for chondritic planets, mass ratio $\gamma=1/6$, and target mass \SI{0.9}{\mearth}. $\xi$ is evaluated using the surrogate model of \citet{2020ApJEmsenhuberA} and differs slightly from the simulations using our updated SPH code. Perfect merger ($\xi\approx1$) is dark blue; the canonical model plots around the bottom center. Increases in velocity or angle (collision angular momentum) lead to a transition from graze-and-merge to hit-and-run ($\xi\approx0$, white). Right: the accretion efficiency of the projectile is computed using the associated surrogate model of \citet{2020ApJEmsenhuberA}, showing fractional mass loss from the surviving runner. The lower-velocity hit-and-runs studied here produce relatively intact runners, $\xi_{\rm imp}\approx 0$.
	}
	\label{fig:phases}
\end{figure*}

The strong tendencies with impact angle and velocity can be appreciated analytically \citep{2020ApJGabriel} and are the basis for physically motivated scaling laws  \citep[e.g.,][]{2012ApJStewart,2016IcarusMovshovitz} and semiempirical models \citep{2010ApJKokubo}. However, detailed 3D numerical simulations are required to assess accretion efficiency in detail.
But while thousands of giant impact simulations have been published at high resolution in 3D, this is still a small sampling of the parameter space, so machine learning has been applied to learn generalized underlying behavior from the published simulations.

\citet{2019ApJCambioni} and \citet{2020ApJEmsenhuberA} obtained surrogate models for accretion efficiency $\xi$ and other giant impact outcomes by training neural networks on 800 high-resolution giant impact simulations for a variety of impact angles and velocities and masses between $0.001-\SI{1}{\mearth}$. The training data \citep{2011PhDReufer,2020ApJGabriel} are high-resolution simulations, about 200,000 SPH particles total, of giant impacts starting from nonrotating bodies of 30/70 wt\% iron/forsterite composition.
A more expansive parameter sampling was generated by \citet{2020CACTimpe}, although at lower resolution (10,000 particles in the projectile). This enabled a more general assessment of machine-learning approaches, applied to a dataset that includes preimpact rotation and compositional variation. They obtained the important result that $\gamma,v_{\rm coll}/v_{\rm esc},\theta_{\rm coll}$ and core mass fraction $Z$ are the most significant parameters in predicting giant impact outcomes. Preimpact rotation is significant, but mostly to the postimpact rotation.

Surrogate models make rapid predictions of known confidence for the outcomes of collisions within the parameter range of the training validation set. They are effectively functions mapping inputs to outputs and thus allow for inversion of evolution models. However, for collisions around the transitions, specific predictions made by surrogate models are less reliable than for major regions (erosion, hit-and-run, merger) because the physical response is nonlinear and the parameter sampling so far is coarse.

The surrogate model for $\xi$ from \citet{2020ApJEmsenhuberA} is plotted in Figure~\ref{fig:phases}, evaluated for $\gamma=1/6$ and for \SI{0.9}{\mearth} targets. Because the training data were generated using the prior version of our code \citep{2011PhDReufer}, this surrogate model computes slightly different outcomes than our updated simulations, but that is not important here.
Note that accretion is more than \SI{50}{\percent} efficient only in the slowest giant impacts, or collisions that are close to head-on. Only a subset of slow collisions have $\xi\sim1$ (dark blue), including the canonical model, which plots near the bottom center.
At somewhat faster velocities the projectile becomes an escaping runner for impact angles greater than \ang{\sim40}, the hit-and-run regime to the right of the graze-and-merge boundary, $\xi\sim0$.
As noted, the plot is for a particular mass ratio, and it would feature more hit-and-run outcomes for larger $\gamma$ and fewer for smaller $\gamma$ for the reasons described above.

We can discriminate the kinds of hit-and-run collisions in terms of the accretion efficiency of the \emph{projectile},
\begin{equation}
\xi_{\rm imp}=(m_{\rm run}-m_{\rm imp})/m_{\rm imp},
\end{equation}
where $m_{\rm run}$ is the mass of the runner. This is analogous to the accretion efficiency of the target (Equation~\ref{eq:xi}), except it is negative because the projectile loses mass. A surrogate model for $\xi_{\rm imp}$ was similarly obtained by training on the same database of simulations \citep{2020ApJEmsenhuberA}, plotted in the right panel of Figure~\ref{fig:phases}, also for $\gamma=$ 1/6.

Any material not in the largest two remnants is called the debris mass $\xi_{\rm esc}$, where by mass conservation $\xi+\xi_{\rm imp}+\xi_{\rm esc}\equiv1$ when debris is measured in units of $m_{\rm imp}$. In a low-velocity hit-and-run, the debris mass is usually much less than one percent of the total colliding mass (\papertwo); the rest is in the postimpact target and runner. In a canonical merger it can be $\sim1-2$ percent (in terms of projectile mass, $\xi\gtrsim\SI{10}{\percent}$), although this is sensitive to the velocity and angle of the collision and the thermal states of the colliding bodies.

At higher velocities and intermediate angles, the escaping projectile from a hit-and-run can be destroyed ($\xi_{\rm imp}<-1/2$) producing voluminous debris even while causing only minor erosion of the target. Multiple massive runners can be derived from one projectile, with highly varied compositions \citep{2006NatureAsphaug,2012PSSSekineGenda}.
Significant erosion of the target requires still higher energy events, in which the projectile escapes as a plume of debris. Relatively head-on impacts are required (upper left; $\xi\ll0$) because otherwise the projectile glances off with little momentum and energy transfer.

Target disruption, a collision removing more than half the target mass, is outside the plot, far upper left. For similar-sized collisions, target disruption requires impact velocity several times $v_{\rm esc}$ and a nearly head-on impact to achieve sufficient coupling. This is faster than can be commonly expected in the late stage, or during any accretionary epoch; otherwise, merger would be rare and too frequently undone.
Catastrophic disruptions of planet-sized targets during the late stage, such as invoked for the origin of Mercury \citep{1988IcarusBenz}, require the scattering influence of migrating gas giants \citep[e.g.,][]{2015ApJCarter} and are then further unusual because of the required small impact angle.

In summary, the simplification that pairwise collisions either result in efficient merger or above some energy threshold destroy the target ignores the vast diversity of outcomes in the middle.
A corollary, once acknowledging that hit-and-runs are common during pairwise accretion, is that even if velocities are too slow for projectiles to destroy their targets, targets frequently destroy projectiles, almost as often as they accrete them, as represented by $\xi_{\rm imp}\leq-0.5$ in Figure~\ref{fig:phases}.

Hit-and-run disruption has been invoked \citep{2007NatureYang} to explain how some iron meteorite parent bodies were disrupted under low-shock conditions and, overall, for the diversity of meteorite progenitors \citep{2017BookAsphaug}.
At planet-forming scales, the loss of Mercury's mantle can be explained by a violent hit-and-run by proto-Mercury into proto-Venus or proto-Earth \citep{2014NatGeoAsphaug,2018ApJChau}, an event that would plot to the left-middle of Figure~\ref{fig:phases}. Given the preservation of volatiles in Mercury \citep[e.g.,][]{2011SciencePeplowski} and the preponderance of lower-speed collisions, the preferred scenario is two nominal-velocity, nominal-impact-angle hit-and-runs in a row \citep{2014NatGeoAsphaug,2018MNRASJackson}---a ``stranded runner'' as demonstrated in \papertwo{}.

The net accretion efficiency in a giant impact increases if the runner or debris eventually get accreted by the target.
The first study to consider the fate of post-Moon-formation debris, by \citet{2012MNRASJacksonWyatt}, modeled the dynamical evolution of escaping material obtained in a canonical simulation by \citet{2009ApJMarcus}. A bit more than one lunar mass escaped ($\xi_{\rm esc}\sim0.13$); of this, they estimated that \SI{17}{\percent} of the debris would end up on Venus and \SI{20}{\percent} on Earth within \SI{10}{\mega\year}, apart from losses to collisional grinding. This would be sufficient to accrete a \SI{>10}{\kilo\meter} layer onto each planet.

In the case of a hit-and-run, the runner may reimpact the target; this too is not uncommon considering the proximity of their orbits. For energetic hit-and-runs with fast, disrupted runners, this can involve a complex chain of subsequent collisions, or no return at all. For lower-energy hit-and-runs, as expected in an accreting population and thus the subject of our study, the runner emerges mostly intact ($\xi_{\rm imp}\sim0$) and escapes more slowly, leading to a higher return probability. These ``hit-and-run return'' collisions are a common mechanism for the late-stage growth of planets (\papertwo).

\subsection{Return to middle-Earth}

For nominal late-stage velocities $v_{\rm coll}\approx 1.2-1.4v_{\rm esc}$, most giant impacts are hit-and-run collisions (Figure~\ref{fig:phases}) that produce relatively intact escaping runners, stripped of a fraction of their atmosphere, hydrosphere, crust, and mantle.
\papertwo{} showed that runners escaping from slow hit-and-runs with proto-Earth return to collide with post-hit-and-run proto-Earth (``middle-Earth'') about half the time, and that an equal fraction go on to collide with (and ultimately accrete with) proto-Venus. Venus loses far fewer of its runners---an asymmetry of formation that is the subject of that paper.

For runners that return to proto-Earth, the interlude between the hit-and-run and the return can be a few thousand years, or longer than $\SI{30}{\mega\year}$.
The average, of order \num{0.1}--\SI{1}{\mega\year}, is much shorter than the error bars in geochronology, so a hit-and-run return origin of the Moon would likely appear geochemically as a single, complex event.

Even if proto-Earth is initially not rotating, the hit-and-run causes it to rotate, although it seldom makes a massive disk. The induced rotation period ($P_{\rm rot}\sim10-11$~hr in the simulations considered below) establishes a preimpact vector for the next collision. A key finding of \paperone{} is that, with the exception of the very early returns ($\lesssim\SI{1000}{\year}$) there is no memory of the orientation from one collision to the next.
The distribution of $\varphi$ is indistinguishable from the expected random distribution $\mathrm{d}p=\frac{1}{2}\sin{\varphi}\mathrm{d}\varphi$, where $\varphi$ is the offset angle between the two collisions. This is true whether or not other planets are included in the integration. Prograde and retrograde returns (parallel or antiparallel, $\varphi=0$ or $180$) are least probable, while the mean and most probable returns are orthogonal, $\varphi=\ang{90}$, the sort that knock a planet on its side.

Because of the transfer of momentum to the target and the loss of kinetic energy to shocks and escaping debris, for nominal impact angles the runner's egress velocity, escaping the postimpact target, is considerably slower than its inbound velocity $v_{\rm coll}$ \citep{2020ApJEmsenhuberA}. Another key finding of \paperone{} is that slower egress velocities lead to comparably slow return velocities.
Given the high likelihood of hit-and-run collisions over the expected velocity range of giant impacts, the slowing down of runners is vital to late-stage planet formation because it leads to the low velocities that are required for accretion. Efficient accretion is not limited to the lower left blue region of the plot, but rather to whatever collisions lead to runners that end up there.

\section{Methods}
\label{sec:methods}

Moon formation by hit-and-run return has three acts that we model independently: the hit-and-run, which slows down proto-Theia and causes initial mixing; the dynamical evolution of Theia's returning orbit; and the canonical-like graze-and-merge collision that forms the Moon.

There are many possibilities for the hit-and-run, and we focus on lower-velocity scenarios that were arguably most common. We sample existing hit-and-run simulations from \paperone{} that start with nonrotating differentiated planets \SI{0.15}{\mearth} and \SI{0.95}{\mearth}, and choose two examples that lie marginally above the hit-and-run transition. For an impact angle of \ang{\sim45} this corresponds to a velocity of $1.15$ to $1.20 v_{\rm esc}$. For these, we calculate the egress velocity of the runner escaping from the target, as well as their postimpact masses and the target rotation. The tendency in this velocity range is for the projectile to be accreted, or to escape relatively intact.

For the second act or interlude, we follow the procedure of \papertwo{} and transfer the target and the escaping runner at their new velocity into the \textit{N}-body code \texttt{mercury} \citep{2012SoftwareChambers}, to track their dynamical evolution. We clone the outcome of each hit-and-run (middle-Earth + runner, ignoring debris) into 1000 random orientations and evolve each cloned set (including the major planets) until there is a follow-on collision, or for \SI{20}{\mega\year} at which time most giant impact chains at \SI{1}{\au} are complete. Longer evolutions up to \SI{400}{\mega\year} are explored systematically in \papertwo{}. From this we obtain the expected distribution of parameters (impact angles and velocities) that define candidate returning-runner scenarios for Moon formation.

The third act is the return collision, which we model using our updated SPH code for several cases within the distribution of return velocities and angles. As there are at least three major parameters to consider, even neglecting rotation and composition, times two collisions, it is not yet possible to explore all cases. For now we focus on the most probable ($\varphi\sim\ang{90},\theta_{\rm coll}\sim\ang{45}$) and fiducial cases, with the intention of hypothesis validation.

We now present the hydrocode methodology, along with the setup of the collisions and the postimpact analysis. We consider various geometries and velocities and include preimpact rotation of the target. We describe the computing of material exchange and the analysis of isotopic mixing in successive collisions. We evaluate the disk mass and the associated angular momentum and composition, applying two different methods in the literature for evaluating the postimpact disk.

\subsection{Giant impact simulations}

Our updated SPH scheme is fully described in \papertwo{}; it is designed and extensively tested for modeling giant impacts (\citealp{2012IcarusReufer,2018IcarusEmsenhuber}; \paperone). General reviews of SPH include \citet{1992ARA&AMonaghan} and \citet{2009NARRosswog}.

SPH is a Lagrangian technique with material distributed into particles. A kernel interpolation is used to compute the quantities at any location, and spatial derivatives are computed using an interpolation with the derivatives of the kernel. To retrieve the neighbor particles for these derivatives, a hierarchical spatial tree is used \citep{1986NatureBarnesHut} that also gives a rapid solution to the self-gravitational potential. In our code, density is retrieved using the kernel interpolation with a correction term for particles close to the surface \citep{2017MNRASReinhardtStadel}.

Entropy that arises as a result of shocks is evolved through artificial viscosity, and pressure is computed from the entropy and density using the (M-)ANEOS EOS \citep{ANEOS,2007M&PSMelosh}. For the terrestrial bodies in this set of papers, we use iron for the core and forsterite Mg$_2$SiO$_4$ (olivine) for the mantle. Friction is ignored for giant impacts of this scale \citep{2018IcarusEmsenhuber}, although it is likely to be important with regard to the structure of debris.

The initialization of pressure, density, and entropy inside each planet is performed in the same way as in \papertwo{}. For spinning planets we also initialize a uniform rotation. First, we retrieve a 1D radial hydrostatic profile using the algorithm of \citet{1991LNPBenz} and the iron/forsterite EOSs. From the 1D profile, an initial SPH 3D spherical body is obtained using the methodology of \citet{2017MNRASReinhardtStadel}.
To establish preimpact rotation, each particle is given a radius-dependent velocity
\begin{equation}
    \vec{v}_\mathrm{ini} = \vec{\Omega} \times \vec{r}_\mathrm{ini}\ ,\ \vec{\Omega}=\left(\begin{array}{c}
         0 \\
         0 \\
         \Omega
    \end{array}\right),
\end{equation}
where $\Omega=2\pi/P$ and $P$ is the desired rotation period. The rotating body is then evolved by itself in SPH with a damping term that uses $\vec{v}_\mathrm{ini}$ as the reference velocity. This forces the body to spin at the prescribed uniform rate and allows the equatorial bulge to form, which changes not only the dynamics but also the collision cross section.

In all simulated collisions, to allow the tidal bulge to form, and for torque to accumulate before the impact, the bodies begin their evolution at a distance of 5 times the sum of the radii, about twice the Roche limit.

\begin{figure*}
	\centering
	\includegraphics{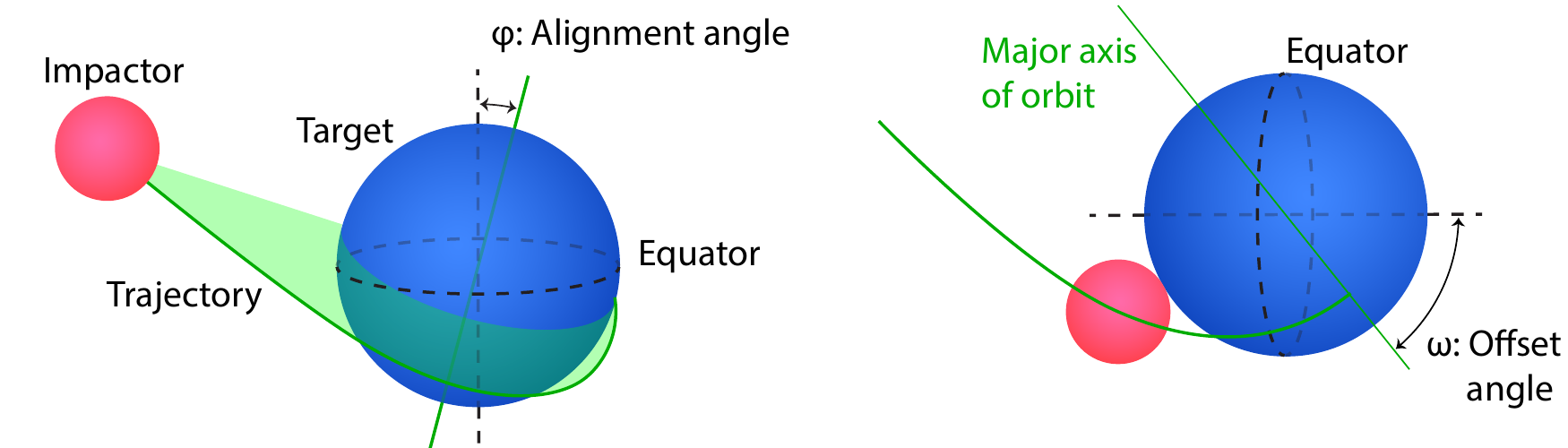}
	\caption{A rotating target has two additional parameters: the alignment angle $\varphi$ between the target equator and the plane of the collision in the two-body frame (left), and the offset angle $\omega$ between the orbit periapse and the spin vector (right). $\varphi$ ranges from prograde (\ang{0}) to retrograde (\ang{180}), the most probable being a perpendicular (\ang{90}) collision.
	The offeset angle (right) is shown for the case $\varphi=\ang{90}$ and an impact pericenter at the pole ($\omega=\ang{0}$). The angular momentum offset $\varphi$ matters most.}
	\label{fig:orientation}
\end{figure*}

\subsection{Return collision setup and rotations}

Bodies from the hit-and-runs are evolved dynamically using the \texttt{mercury} code \citep{2012SoftwareChambers} for 1000 clones representing the velocity $\sim1.01v_{\rm esc}$ in a distribution of directions. As we shall see, the return collision ends up being $v_{\rm coll}\lesssim{}1.05 v_{\rm esc}$ in most cases, so we model return collisions as either $1.00$ or $1.05v_{\rm esc}$, fixing the impact angle at \ang{\sim45} in each case.

The targets must be set up with a rotation period $P_{\rm rot}=$10-11~hr, the result of the hit-and-runs described. This requires specification of the orientation of the return collision (Figure~\ref{fig:orientation}). The alignment angle $\varphi$ is between the rotation axis of the target and the orbital angular momentum vector in the two-body frame of the collision, while the offset angle $\omega$ is between the periapse (as if there were no collision) and target spin vector.
For now we apply zero rotation to the reimpacting runner, although barely escaping runners tend to rotate rapidly \citep{2006NatureAsphaug}, which this might be important to disk composition.

For random return collisions it was shown in \paperone{} that the offset angle $\omega$ is uniformly distributed, while $\varphi$ follows the probability distribution $\mathrm{d}p=\frac{1}{2}\sin{\varphi}\mathrm{d}\varphi$ with a maximum at $\varphi=\ang{90}$.
For the limited set of studies presented here, we therefore focus on impact angles of around \ang{45}, and on alignment angles of around \ang{90}, with other alignment angles (prograde, retrograde, and nonrotating) for comparison.
As with the first collision, we start the bodies at a distance of 5 times the sum of the radii.

\subsection{Determination of the protolunar disk}

In our proposed scenario the protolunar disk is a consequence of the terminal collision, when the runner's mass and angular momentum are finally accreted. A slow hit-and-run usually does not produce a massive disk (\papertwo{}), but it has three primary effects: it spins up the target ahead of the next collision, it exchanges an initial mass fraction between the colliding bodies (middle panel of Figure~\ref{fig:schema}), and it causes significant heating of the target that might influence the terminal collision.

After the giant impact has evolved in SPH to the point that there are no further major collisional interactions, we evaluate the disk. (By ``disk'' we mean any postimpact orbiting material, despite the possibility of the mass being dominated by huge clumps.) We first identify all particles that are gravitationally bound to the single largest remnant; escaping particles are removed from the analysis.
The surface of the largest remnant is then computed as the layer whose density is approximately three-quarters of the reference density of the EOS. The nonescaping particles outside that surface are potential disk particles that are then evaluated according to two approaches.

The first approach (\citealp{2012IcarusReufer} as adapted by \citealp{2019ApJEmsenhuberA}) is to calculate the pericenter of the two-body problem for each potential disk particle about the central body. If the pericenter is below the identified surface, the particle is added to Earth; otherwise, it is added to the disk. The second approach \citep[e.g.,][]{2001IcarusCanup,2004IcarusCanup} is to calculate the radius of a circular orbit with the equivalent angular momentum to each potential disk particle and then proceed similarly, adding the particle to the disk if this equivalent circular orbit is outside the identified surface. The latter is more inclusive, as all particles having their pericenter above the surface also have their equivalent radius above the surface. We include both approaches in our analysis to compare with published results for post-impact disks.

For the case of a fluid-particle postimpact disk, the final coagulation to form the Moon is at best \SI{\sim50}{\percent} efficient according to \citet{2012ApJSalmon}, consistent with the results of \citet{1997NatureIda}. If such disk approximations are correct, then to be successful a giant impact needs to place at least 2 lunar masses into orbit, although this depends on disk angular momentum \citep{2000IcarusKokuboB}.

\subsection{Material exchange in two collisions}

Two collisions in a row provide two stages of material exchange, as illustrated in Figure~\ref{fig:schema}. The fractions of proto-Theia and proto-Earth in the final Earth's mantle and the protolunar disk are computed by convolving the material exchanges/contributions calculated in both collisions individually. We assume that complete mixing occurs within each silicate reservoir during the interlude and that the bodies are fully differentiated.
Core material is not exchanged between colliding bodies in these relatively low energy events.
We have not attempted to evaluate any metal-silicate equilibration that may occur inside each body response to either collision. While equilibration could have significant implications, for example, in the resetting of $^{182}$W \citep[see][]{2015IcarusDwyer}, such a study is outside the scope of our analysis, which focuses on silicate compositions.

The first stage of silicate mass exchange is defined by the equilibration factor \citep{2012IcarusReufer}
\begin{equation}
    \delta f_\mathrm{T}^1=\frac{f^\mathrm{s\leftarrow t}_\mathrm{mant}}{f^\mathrm{l \leftarrow t}_\mathrm{mant}}-1,
\end{equation}
where $t$ and $i$ stand for the original target  (proto-Earth) and impactor (proto-Theia) of the hit-and-run, respectively, and $l$ and $s$ stand for the largest remnant (middle-Earth) and second-largest remnant (Theia, the runner) of that collision, respectively. Thus, $f^\mathrm{s\leftarrow t}_\mathrm{mant}$ is the mantle mass fraction of Theia that comes from proto-Earth, $f^\mathrm{l\leftarrow t}_\mathrm{mant}=1-f^\mathrm{l\leftarrow i}_\mathrm{mant}$ is the mantle mass fraction of middle-Earth that comes from proto-Earth, and $f^\mathrm{l\leftarrow i}_\mathrm{mant}$ is the mantle mass fraction of middle-Earth that comes from the impactor. These mixing values are given in Table~1 of \papertwo{} for a range of giant impacts, from which we select our starting hit-and-run collisions.

The combined silicate equilibration factor is computed as the product of both collisions,
\begin{equation}
    \delta f_\mathrm{T}^\mathrm{c}=\frac{f^\mathrm{d \leftarrow t}_\mathrm{mant}}{f^\mathrm{e \leftarrow t}_\mathrm{mant}}-1,
\end{equation}
where the superscript ``$\mathrm{d \leftarrow t}$'' indicates the fraction of the protolunar disk that originates from the proto-Earth, and superscript ``$\mathrm{e \leftarrow t}$'' indicates the fraction of Earth that does.
These give the cumulative relative contributions in the mantles of the Moon and Earth.
Under the assumption of complete mixing within each silicate reservoir during the interlude, we can then write
\begin{eqnarray}
f^\mathrm{d \leftarrow t}_\mathrm{mant} & = & f^\mathrm{d \leftarrow l}_\mathrm{mant} f^\mathrm{l \leftarrow t}_\mathrm{mant} + (1-f^\mathrm{d \leftarrow l}_\mathrm{mant}) f^\mathrm{s \leftarrow t}_\mathrm{mant}\ \ \  \mathrm{and} \\
f^\mathrm{e \leftarrow t}_\mathrm{mant} & = & f^\mathrm{e \leftarrow l}_\mathrm{mant} f^\mathrm{l \leftarrow t}_\mathrm{mant} + (1-f^\mathrm{e \leftarrow l}_\mathrm{mant}) f^\mathrm{s \leftarrow t}_\mathrm{mant}.
\end{eqnarray}
Putting everything together results in a final combined mantle equilibration ratio
\begin{equation}
    \delta f_\mathrm{T}^\mathrm{c} = \frac{1+\delta f_\mathrm{T}^1(1-f^\mathrm{d \leftarrow l}_\mathrm{mant})}{1+\delta f_\mathrm{T}^1(1-f^\mathrm{e \leftarrow l}_\mathrm{mant})}-1.
    \label{eq:dftc-eq}
\end{equation}

If $f_\mathrm{T}^1=-1$, that is, no mixing in the hit-and-run, then $\delta f_\mathrm{T}^\mathrm{c}$ is equivalent to
\begin{equation}
    \delta f_\mathrm{T}^2=\frac{f^\mathrm{d \leftarrow l}_\mathrm{mant}}{f^\mathrm{e \leftarrow l}_\mathrm{mant}}-1,
    \label{eq:dft2}
\end{equation}
which, as expected, is the equilibration of the second collision alone.
A further intuitive understanding of Equation~(\ref{eq:dftc-eq}) is obtained by assuming $f^\mathrm{e \leftarrow l}_\mathrm{mant}=1$ and using Equation~(\ref{eq:dft2}) to replace $f^\mathrm{d \leftarrow l}_\mathrm{mant}$ so that all of middle-Earth's mantle is made only of proto-Earth mantle. Then,
\begin{equation}
     \delta f_\mathrm{T}^\mathrm{c}\approx -\delta f_\mathrm{T}^1 \delta f_\mathrm{T}^2.
\end{equation}
Across our simulations, we obtain $f^\mathrm{e \leftarrow l}_\mathrm{mant}\approx\num{0.96}$, so this assumption results in reasonable values for the estimation of $\delta f_\mathrm{T}^\mathrm{c}$.

\section{Results}

\subsection{Constraining the first collision}

The first giant impact is defined by our hypothesis to be a hit-and-run that meets three basic requirements: the runner is approximately \SI{0.1}{\mearth} per the canonical model; the runner retains a massive silicate mantle, i.e., is more or less chondritic; and the egress velocity is slow so that it is likely to be followed by a low-velocity return. This implies a proto-Theia with a mass of order \SI{0.15}{\mearth}, and a hit-and-run collision velocity $v_{\rm coll}\lesssim{1.2}v_{\rm esc}$, depending on angle.

We do not consider impacts that are close to head-on, in this velocity range, because these end up being accretions that are incompatible with Moon formation, having insufficient angular momentum.
As for highly grazing initial impacts, these have much less direct mass intersection, so the runner is not sufficiently slowed down and emerges largely intact, although not without significant geophysical transformations \citep{2006NatureAsphaug}. From the point of view of Moon formation a highly grazing collision may serve primarily as a dynamical scattering event, without significant mass exchange or deceleration, so we do not consider these explicitly either in our current study.

We have also not explicitly considered giant impacts faster than $1.2 v_{\rm esc}$. In close to head-on cases these can also be accretions, though not efficient, and incompatible with Moon formation. At higher velocity they can be disruptive to the projectile and erosive to the target. At more typical impact angles (\ang{\sim30}-\ang{60}) these faster impacts are hit-and-runs causing significant mass loss from the projectile ($\xi_{\rm imp}\ll0$), or projectile disruption, and erosion of the target.
We have not attempted to model the latter, simply because the egress velocity from a fast hit-and-run is usually also fast, so that the return collision is not as likely. When it occurs, it is itself likely to be a hit-and-run, if not a head-on merger. A faster hit-and-run could therefore be the start of a longer chain, explored in our previous papers, in which case our scenario could be the end of a series of collisions, in a system more laden with debris.

Focusing on impact angles \ang{\sim45}, target masses \SI{0.9}{\mearth}, impactor masses \SI{0.15}{\mearth} ($\gamma=1/6$), and chondritic compositions, two cases from \papertwo{} are just above the hit-and-run threshold, one at $1.15v_{\rm esc}$ at \ang{47.5}, and another at $1.20v_{\rm esc}$ collision at \ang{43}. The parameters and outcomes of these collisions are given in Table~\ref{tab:hrc} and are indicated by the ``Prior'' column in Table~\ref{tab:moon}.
The slower, more grazing hit-and-run produces a somewhat more massive runner that is ultimately more successful at protolunar disk production. A larger runner could of course be obtained in the faster case by starting with a larger proto-Theia; these two specified cases are intended as examples.
The core mass fraction $Z_{sr}$ of the runner following each collision is somewhat greater than the starting value $Z_{\rm imp}=\SI{30}{\percent}$, due to mantle stripping, which is more significant in the faster, somewhat more head-on collision.

All simulations start with nonrotating proto-Earths and proto-Theias. The largest remnant (middle-Earth, subscript ``l'' in Equation~\ref{eq:dft2}) ends with a rotation period $\sim11$ hours following the slower collision and $\sim10$ hours after the faster, more head-on collision.
According to \citet{2008IcarusCanup}, this is where preimpact spin begins to play an important role in giant impacts, so we include it in setting up the return collision.

The deflection angles in these hit-and-runs, relative to the incoming vector, are \ang{52} and \ang{57}, measured between the approach vector and the egress vector, where the larger deflection is for the egress that just barely escapes, the faster impact. If the approach angle is uniformly distributed in the center-of-mass frame of the collision, then the deflection angle does not play an important role.

So the four principal outcomes of the hit-and-run, in order of significance to Moon formation, are deceleration of the runner, spin-up of the target, heating of the target, and mantle (silicate) loss from the projectile. A slow hit-and-run usually does not produce a massive disk.

\begin{table}
    \centering
    \caption{The hit-and-run slows down proto-Theia from a starting value $1.15-2.0 v_{\rm esc}$ to a runner egress velocity $\sim1.01 v_{\rm esc}$ (Figure~\ref{fig:return-velocity}).}
    \begin{tabular}{cccccccc}
        \hline
        $\frac{v_\mathrm{coll}}{v_\mathrm{esc}}$ & $\theta_\mathrm{coll}$ & $m_\mathrm{lr}$ & $m_\mathrm{sr}$ & $P_\mathrm{lr}$ & $Z_\mathrm{lr}$ & $Z_\mathrm{sr}$ & $\delta f_\mathrm{T}^1$ \\
        & [deg] & [\si{\mearth}] & [\si{\mearth}] & [hr] & & & \\
        \hline
        \num{1.15} & \num{47.5} & \num{0.94} & \num{0.11} & \num{11} & \SI{30}{\percent} & \SI{32}{\percent} & \SI{-83.6}{\percent} \\
        \num{1.20} & \num{43.0} & \num{0.95} & \num{0.081} & \num{10} & \SI{30}{\percent} & \SI{36}{\percent} & \SI{-75.4}{\percent} \\
        \hline
    \end{tabular}
    \tablecomments{It removes some mantle from proto-Theia, reducing its mass and increasing its iron fraction ($Z_\mathrm{sr}$) by several percent. It does not make a disk, but spins up the largest remnant which is important to the follow-on collision.}
    \label{tab:hrc}
\end{table}

\subsection{Dynamical evolution of the runner}

A slow egress velocity from a hit-and-run collision will usually lead to a correspondingly slow return collision, which is likely to be a graze-and-merge collision for most impact angles, thus compatible with Moon formation. However, this depends on the dynamical perturbations the runner has along the way. The longer the interlude between the collisions, the longer the opportunity for encounters with proto-Earth, proto-Venus, or other bodies that would energize the subsequent encounters.

\begin{figure}
    \centering
    \includegraphics{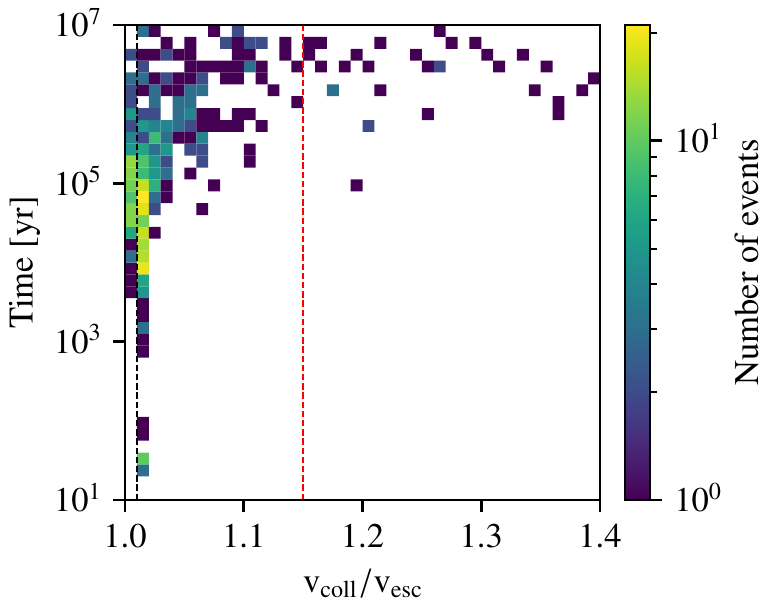}
    \caption{
    2D histogram of impact velocity (normalized to $v_{\rm esc}$) versus time between the collisions, for hit-and-run returns with proto-Earth at 1~AU (from Table~1 in \papertwo{}). Dynamics are computed including the major Solar System planets starting on their current orbits.
    Plotted is the slower hit and run, $v_\mathrm{coll}/v_\mathrm{esc}=1.15$
    (red dashed line), for parameters
    $\theta=\ang{47.5}$,
    $m_\mathrm{imp}=\SI{0.15}{\mearth}$ and
    $m_\mathrm{tar}=\SI{0.90}{\mearth}$.
    The escaping runner velocity is $v_\mathrm{run}/v_\mathrm{esc}=1.01$ (black dashed line).
    }
    \label{fig:return-velocity}
\end{figure}

This tendency is shown in Figure \ref{fig:return-velocity}, which summarizes the dynamical evolution of 1000 clones of the middle-Earth and Theia (remnants of each hit-and-run, integrated with the present major planets) for egress velocities $1.01v_{\rm esc}$ in the center-of-mass frame starting at \SI{1}{\au}. Plotted is a 2D histogram of the return velocity ($v_{\rm imp}/v_{\rm esc}$) versus number of years between the two giant impacts, for the distribution of clones launched in random directions. The hit-and-run velocity $v_{\rm imp}/v_{\rm esc}=1.15$ is shown by the red dashed line, and the egress velocity $v_{\rm run}/v_{\rm esc}=1.01$ is shown by the black dashed line. Histogram values are logarithmic, with yellow indicating the majority of returns.

Return collisions happen over a range of times, mostly within $\sim 10^4$--$10^6$ years. Early returns are close to the egress velocity, which is close to $v_{\rm esc}$; later returns happen at a wider range of velocities owing to gravitational encounters. Nonimpacting runners are not plotted in Figure \ref{fig:return-velocity} but are studied in \papertwo{}; these either move on to another planet (usually Venus) or become part of the background population.

\subsection{Final accretion and Moon formation}
\label{sec:res-acc}

We simulate the return collisions using the SPH methods described in Section~\ref{sec:methods} and \papertwo{}. It is not feasible to span the possibilities of mass ratio, velocity, core size, collision angle, and rotation vector, so we limit our current analysis to 10 cases that seem representative of the predicted terminal collisions, based on largest remnants from two of the hit-and-run collisions in \papertwo{}, plus a simulation of the canonical case for comparison.

We use the largest remnants of the hit-and-runs in Table~\ref{tab:hrc} to prescribe targets of the appropriate mass and rotation period and projectiles (runners) of the appropriate mass.
The target of each return collision is set at \SI{\sim0.95}{\mearth} in each case. The projectile is \SI{0.11}{\mearth} in the slower case and \SI{0.08}{\mearth} in the faster case. The alignment angle (Figure~\ref{fig:orientation}) ranges from prograde ($\varphi=0$) to retrograde ($\varphi=\ang{180}$), the most likely being \ang{90}, so we focus on orthogonal collisions. We also vary the offset angle $\omega$ although this is less significant.
The same composition and entropy profiles are applied as in the original collision, and target rotations are applied as described in Section~\ref{sec:methods}. As discussed further below, a much hotter middle-Earth may lead to a more massive, more Earth-composition disk \citep{2019NatGeoHosono}, but we do not consider this currently.

Each collision in Table~\ref{tab:moon} was evolved to at least \SI{48}{\hour} after contact, in order for graze-and-merge collisions to complete and for the reported disk quantities to converge. In two cases at slightly more grazing incidence (\ang{48}, discussed further below) the integration time was doubled to follow their extended graze-and-merge orbits. One of the challenges of giant impact simulations in the accretion regime is that many scenarios can require days to evolve the point that an outcome (merger vs. hit-and-run) is determined. This can require significant machine time, plus greater long-range dynamical accuracy than SPH provides. It also makes it difficult to compare outcomes for disk mass and angular momentum, when disk states are evaluated at different times.

Results are given in Table~\ref{tab:moon} for the total escaping mass not bound to the final system, the total bound angular momentum, and the mass, angular momentum, inclination, and composition of the disk. Two sets of results are given corresponding to the two approaches (see Section~\ref{sec:methods}) for determining what particles belong to the disk at the end of a simulation. Overall, for a number of our modeled return collisions the final properties are comparable to those obtained in canonical simulations, with an improvement in most cases, especially with regard to isotopic mixing. We have not performed any tuning to obtain a favorable result, other than to consider hit-and-runs that end with a low-velocity runner and a return impact angle close to the expected value.

\begin{table*}
\movetabledown=2in
\begin{rotatetable*}
    \centering
    \caption{Results from Our SPH Simulations of the Terminal Accretion Collisions}
    \begin{tabular}{|ccccc|cc|cccccc|cccccc|}
        \hline
        & & & & & & & \multicolumn{6}{c|}{Pericenter above surface} & \multicolumn{6}{c|}{Equivalent radius above surface} \\
        \cline{8-13} \cline{14-19}
        Prior & $\frac{v_\mathrm{coll}}{v_\mathrm{esc}}$ & $\theta_\mathrm{coll}$ & $\varphi$ & $\omega$ & $m_\mathrm{esc}$ & $L_\mathrm{tot}$ & $m_\mathrm{disk}$ & $L_\mathrm{disk}$ & $\frac{m_\mathrm{Fe}}{m_\mathrm{disk}}$ & $\phi_\mathrm{disk}$ & $\delta f_\mathrm{T}^2$ & $\delta f_\mathrm{T}^\mathrm{c}$ & $m_\mathrm{disk}$ & $L_\mathrm{disk}$ & $\frac{m_\mathrm{Fe}}{m_\mathrm{disk}}$ & $\phi_\mathrm{disk}$ & $\delta f_\mathrm{T}^2$ & $\delta f_\mathrm{T}^\mathrm{c}$ \\
        & & [deg] & [deg] & [deg] & [\si{\mmoon}] & [\si{\lem}] & [\si{\mmoon}] & [\si{\lem}] & & [deg] & & & [\si{\mmoon}] & [\si{\lem}] & & [deg] & & \\
        \hline
        c. case*    & \num{1.00} & \num{45} & --        & --       & \num{0.175} & \num{0.950} & \num{0.887} & \num{0.168} & \SI{1.0}{\percent} & \num{0.3} & \SI{-58.5}{\percent} & -- & \num{0.964} & \num{0.173} & \SI{3.6}{\percent} & \num{0.3} & \SI{-57.2}{\percent} & -- \\
        \num{1.15}* & \num{1.00} & \num{45} & --       & --       & \num{0.079} & \num{1.100} & \num{1.120} & \num{0.224} & \SI{1.6}{\percent} & \num{0.3} & \SI{-56.8}{\percent} & \SI{-47.2}{\percent} & \num{1.218} & \num{0.229} & \SI{4.1}{\percent} & \num{0.2} & \SI{-56.1}{\percent} & \SI{-46.1}{\percent} \\
        \num{1.15} & \num{1.00} & \num{45} & \num{0}   & --       & \num{0.113} & \num{1.430} & \num{1.433} & \num{0.295} & \SI{1.4}{\percent} & \num{0.3} & \SI{-62.4}{\percent} & \SI{-51.8}{\percent} & \num{1.547} & \num{0.298} & \SI{3.6}{\percent} & \num{0.4} & \SI{-62.2}{\percent} & \SI{-51.2}{\percent} \\
        \num{1.15} & \num{1.00} & \num{45} & \num{180} & --       & \num{0.160} & \num{0.744} & \num{0.745} & \num{0.145} & \SI{0.5}{\percent} & \num{0.9} & \SI{-56.1}{\percent} & \SI{-46.6}{\percent} & \num{1.003} & \num{0.154} & \SI{7.9}{\percent} & \num{0.9} & \SI{-57.4}{\percent} & \SI{-47.2}{\percent} \\
        \num{1.15} & \num{1.00} & \num{45} & \num{90}  & \num{0}  & \num{0.081} & \num{1.152} & \num{1.101} & \num{0.219} & \SI{0.7}{\percent} & \num{17.5} & \SI{-58.4}{\percent} & \SI{-48.5}{\percent} & \num{1.224} & \num{0.224} & \SI{4.8}{\percent} & \num{17.5} & \SI{-57.9}{\percent} & \SI{-47.6}{\percent} \\
        \num{1.15} & \num{1.00} & \num{45} & \num{90}  & \num{90} & \num{0.194} & \num{1.123} & \num{1.159} & \num{0.225} & \SI{2.1}{\percent} & \num{20.1} & \SI{-57.0}{\percent} & \SI{-47.4}{\percent} & \num{1.307} & \num{0.230} & \SI{8.2}{\percent} & \num{20.0} & \SI{-56.2}{\percent} & \SI{-46.3}{\percent} \\
        \num{1.15} & \num{1.00} & \num{48} & \num{90}  & \num{0}  & \num{0.062} & \num{1.208} & \num{1.403} & \num{0.331} & \SI{0.1}{\percent} & \num{19.8} & \SI{-73.1}{\percent} & \SI{-60.8}{\percent} & \num{2.131} & \num{0.338} & \SI{27.8}{\percent} & \num{19.6} & \SI{-73.8}{\percent} & \SI{-60.8}{\percent} \\
        \num{1.15} & \num{1.05} & \num{45} & \num{90}  & \num{0}  & \num{0.994} & \num{1.049} & \num{0.834} & \num{0.173} & \SI{2.3}{\percent} & \num{19.4} & \SI{-52.0}{\percent} & \SI{-43.2}{\percent} & \num{1.569} & \num{0.178} & \SI{42.4}{\percent} & \num{18.9} & \SI{-52.3}{\percent} & \SI{-43.1}{\percent} \\
        \num{1.15} & \num{1.05} & \num{48} & \num{90}  & \num{0}  & \num{0.356} & \num{1.214} & \num{1.283} & \num{0.308} & \SI{0.1}{\percent} & \num{16.6} & \SI{-75.5}{\percent} & \SI{-62.9}{\percent} & \num{2.870} & \num{0.325} & \SI{44.0}{\percent} & \num{15.7} & \SI{-77.5}{\percent} & \SI{-63.9}{\percent} \\
        \num{1.20} & \num{1.00} & \num{45} & \num{0}   & --       & \num{0.092} & \num{1.172} & \num{0.812} & \num{0.146} & \SI{1.4}{\percent} & \num{1.6} & \SI{-67.3}{\percent} & \SI{-50.5}{\percent} & \num{0.858} & \num{0.148} & \SI{1.8}{\percent} & \num{1.6} & \SI{-66.3}{\percent} & \SI{-49.2}{\percent} \\
        \num{1.20} & \num{1.00} & \num{45} & \num{90}  & \num{0}  & \num{0.506} & \num{0.828} & \num{0.292} & \num{0.054} & \SI{0.1}{\percent} & \num{23.2} & \SI{-50.5}{\percent} & \SI{-37.9}{\percent} & \num{0.352} & \num{0.055} & \SI{6.8}{\percent} & \num{22.9} & \SI{-50.0}{\percent} & \SI{-37.7}{\percent} \\
        \hline
    \end{tabular}
    \tablecomments{The first five columns are the initial conditions, with ``prior'' defining the body properties from the corresponding entries in Table~\ref{tab:hrc}. The first row is our simulation of the canonical case (\SI{0.9}{\mearth} target, \SI{0.1}{\mearth} projetile, \SI{30}{\percent} core mass fraction) and the star (*) is for non-rotating targets. The total escaping mass $m_\mathrm{esc}$ and total angular momentum $L_\mathrm{tot}$ are shown, and then the disk properties computed using both approaches (see text): disk mass $m_\mathrm{disk}$, disk angular momentum $L_\mathrm{disk}$, disk iron mass fraction $\frac{m_\mathrm{Fe}}{m_\mathrm{disk}}$, disk inclination $\phi_\mathrm{disk}$, and final disk-planet compositional imbalance $\delta f_\mathrm{T}^\mathrm{c}$.}
    \label{tab:moon}
\end{rotatetable*}
\end{table*}

\subsubsection{Protolunar disk mass, angular momentum, and composition}

SPH simulations of giant impacts, and especially their disks, become inaccurate after tens of hours for the reasons described above, yet the postimpact state continues to evolve dynamically. The disk quantities must therefore be estimated, which is done by treating each nonescaping SPH particle as a test body that orbits the largest remnant and applying a criterion, as was described in Section~\ref{sec:methods}. The more conservative criterion \citep[e.g.,][]{2000IcarusKokuboB,2012IcarusReufer} is that the pericenter of each particle's orbit must be outside the computed target surface (`pericenter above surface' in Table~\ref{tab:moon}). The other approach \citep[e.g.,][]{2001IcarusCanup,2004IcarusCanup} is to convert each SPH particle's orbit into a circular orbit with the equivalent angular momentum. Here we define the equivalent radius as a circular orbit of semimajor axis $a(1-e^2)$, where $e$ is the eccentricity of the particle orbit. If this circular orbit is outside the target, the particle it is added to the disk (`equivalent radius above surface' in Table~\ref{tab:moon}).

The mass and composition of the postimpact orbiting material (not always disk-like) are found in simulations \citep[e.g.,][]{2013IcarusCanup} to evolve rapidly owing to the reimpact of massive projectile-remnant clumps. This is apparent when comparing some of the pericenter and equivalent-radius disk evaluations applied to our simulations. Three of the cases end up with several-\si{\mmoon} disks in the equivalent-radius evaluation (\num{1.6} to \SI{2.9}{\mmoon}), but their iron mass fractions are large, \SI{28}{\percent} to \SI{44}{\percent}, respectively. In the pericenter evaluation of the same SPH outcomes, these cases end up with less than \SI{2}{\percent} iron in the disk, and at most \SI{1.3}{\mmoon} disk mass. This is because iron-rich clumps come crashing down in the stricter pericenter evaluation.

Using the pericenter evaluation, we find that hit-and-run return collisions can lead to silicate disks significantly greater than one lunar mass, in a final system with the right $L_{\rm EM}$, and with a low iron mass fraction.
The angular momentum in the postimpact disk, expressed as $L_{\rm disk}/L_{\rm EM}$ in Table~\ref{tab:moon}, equals or exceeds the proposed minimum value of $\sim0.18$ \citep[e.g.,][]{2001IcarusCanup} in most of the simulated cases. The disk mass and disk angular momentum are both greater, in most of our return collisions, than in our simulation of the canonical model.

When using the equivalent-radius evaluation for these same simulations, the disk masses are substantially greater, although the disk angular momentum only slightly increases. This is because this criterion includes higher-eccentricity particles that are rejected by the pericenter evaluation; the equivalent radius of these particles is only slightly above the surface. Using the scaling of \citet{2000BookKokubo}, a disk with a small specific angular momentum is unable to coalesce into a large satellite; material with a specific angular momentum lower than that of an equivalent orbit at $\approx 0.6 R_\mathrm{Roche}=1.7R_\oplus$ does not contribute to the final satellite. Disk mass estimates obtained with the equivalent-radius approach are thus not representative of the final mass of the satellite.

When comparing equivalent methods of evaluating the postimpact disk, we see that hit-and-run return collisions can lead to solutions comparable to, or improving upon, simulations of the canonical case. This is also true for disk isotopic composition. The final protolunar disk is composed of proto-Theia and proto-Earth materials, with proportions obtained over the course of two collisions using the conventions of \citet{2012IcarusReufer} and \papertwo{} (Equation~\ref{eq:dftc-eq}). This is compared to the mantle composition of the final target, also derived from both components in two collisions, to compute the final difference, Moon vs. Earth composition, $f_\mathrm{T}^c$ in Table \ref{tab:moon}, which would be $0$ for equally derived composition, and \SI{-100}{\percent} for lunar silicates derived entirely from proto-Theia.
Compositional similarity is improved by $\gtrsim\SI{10}{\percent}$ compared to the canonical case, depending on the use of the pericenter or equivalent-radius criterion for which particles end up in the disk. The equivalent-radius approach also tends to leave more of Theia's iron in the disk.

The improvement in isotopic equilibration, although significant, is not sufficient to erase any major differences in proto-Earth/proto-Theia composition. As discussed above, their similarity might not have to be so exact \citep[e.g.,][]{2020NatGeoCano}. But we consider this aspect of our results further, noting that our computed $f_\mathrm{T}^c$ is likely to underestimate the final homogenization for several reasons.

For one thing, SPH tends to suppress mixing at interacting boundaries \citep[e.g.,][]{2019ApJDengB}. This might artificially limit the material exchange in the hit-and-run collision when the planets shear past and through one another. Hit-and-run is not a ``bounce'' but an intense thermophysical and dynamical contact lasting about half an hour.
Suppression of mixing might also limit the entrainment of Earth materials into the disk, which means that canonical scenarios might improve as well, if mixing in greater than computed. But the effect would be of compounded significance in collision-chain scenarios.

For another, Theia and middle-Earth are initialized, in our setup for the terminal collision, with the same subsolidus entropy profile as proto-Theia and proto-Earth in the first collision (see Section~\ref{sec:methods}). Although hit-and-runs do not impart the same degree of heating as accretions, for the reasons noted above, they are nevertheless violently energetic and can produce a global magma ocean \citep{2021EPSLNakajima}. If so, and the second collision happened before it cooled, then according to \citet{2019NatGeoHosono} this would enhance protolunar disk production and increase the fractional contribution from proto-Earth.

Furthermore, our simplified representation of hit-and-run as two bodies coming in and two bodies going out, like a bounce, ignores the fact that a collision chain generates substantial heliocentric debris (\papertwo{}). Much of this will return within a few million years \citep[i.e.,][]{2012MNRASJacksonWyatt} while Earth and the Moon are closely orbiting, potentially leading to ballistic material exchanges and other modifications. Debris in a hit-and-run can be tens of percent of the total colliding mass, depending on starting velocity (Figure~\ref{fig:phases}), but for slow collision chains debris production is dominated by the final merger ($m_{\rm esc}$ in Table~\ref{tab:moon}). Faster, more debris-laden giant impacts are relatively common in terrestrial planet formation but have yet to be modeled in the context of our proposed scenario.

\subsubsection{Sensitivity to parameters}

There is considerable sensitivity to the return impact angle and velocity, even within our very limited range of collisional parameters. Four of the modeled return collisions are set up with $\varphi=\ang{90}$ and $P_{\rm rot}=11$\SI{}{\hour} and the same colliding bodies, only varying the impact angle, $\theta_\mathrm{coll}=\ang{45}$ or \ang{48}, and velocity, $v_\mathrm{coll}=1.00$ or $1.05 v_\mathrm{esc}$.
The $\theta_\mathrm{coll}=\ang{45}$ collisions are graze-and-merge events ending in accretion, resembling the canonical case but with pronounced disk inclination. The $\theta_\mathrm{coll}=\ang{48}$ collisions are also graze-and-merge, but the bound runner misses the target when it loops back, passing inside the threshold for tidal mass loss \citep{1992IcarusSridharTremaine} so that the runner's energy is further reduced.

In the $1.00v_{\rm esc}$ case at \ang{48}, this leads to subsequent accretion and the spin-out of a massive disk.
For the $1.05v_{\rm esc}$ case we end the simulation after $t=\SI{96}{\hour}$, by which time dissipation has begun to circularize the runner's orbit with a period of \SI{\sim4}{\hour} around the target. It is unclear whether the bodies will merge, as in the previous case, or end in a direct capture reminiscent of early models of lunar formation \citep{1986IcarusBenz} and scenarios for Pluto-Charon formation \citep{2011AJCanup}. Because the runner orbits just inside the corotation radius of Earth, it is expected to eventually collide, and could then make a delayed protolunar disk.

As for prograde collisions ($\varphi=0$) into spinning targets, these are found to produce more massive disks, and greater system angular momentum overall, but at the expense of resulting in somewhat less mixed isotopic compositions (e.g., $L_\mathrm{tot}=\SI{1.433}{\lem}$ and $\delta f_\mathrm{T}^c=\SI{-51.8}{\percent}$ for the case with prior $v_\mathrm{coll}/v_\mathrm{esc}=\num{1.15}$ and $\varphi=\ang{0}$).

Retrograde collisions lead to insufficient disk mass and lower combined angular momentum in our study. For example, $m_\mathrm{disk}=\SI{0.745}{\mmoon}$ and $L_\mathrm{tot}=\SI{0.744}{\lem}$ for the case with prior $v_\mathrm{coll}/v_\mathrm{esc}=\num{1.15}$ and $\varphi=\ang{180}$.
Moon formation might still be possible in retrograde collisions, but with a more massive, higher-velocity projectile providing larger orbital angular momentum to compensate for the opposing spin.

For the retrograde collisions we have studied, the angular momentum acquired is sufficient to reverse the rotation of the counterrotating proto-Earth, so the disk (although less than a lunar mass in these cases) and target end up rotating in the same direction. The Earth's rotation ends up being slower by about a factor of two than for the prograde but otherwise-identical collision, \SI{8.9}{\hour} versus \SI{4.5}{\hour}. We do not recover the trend identified by \citet{2008IcarusCanup}, where the escaping mass $m_\mathrm{esc}$ is larger for retrograde than for prograde impacts.

\subsubsection{Target rotation and disk inclination}

To first order, we find that protolunar disks produced in orthogonal (that is, expected) cases are similar to nonrotating canonical cases: silicate disks of $\sim1-2$ lunar masses with very little iron. But there are two important differences: impacts into rotating targets liberate more material into heliocentric orbit, likely due to the lower preimpact binding energy of the rotating target, and the disk ends up highly inclined to the rotating postimpact Earth in the case of an orthogonal collision.

In prograde collisions we find that the disk ends up aligned within about \ang{1} to the equator of Earth. But in orthogonal collisions the postimpact disks end up inclined by up to $\sim\ang{20}$ in the cases we have modeled. This is because the disk is formed almost entirely in the return collision and thus retains Theia's angular momentum, while the Earth's final spin is the combination of the hit-and-run and the terminal accretion.

A strongly inclined protolunar disk is an intriguing outcome of our model, given the Moon's pronounced and unexplained orbital inclination with respect to the Earth's equator \citep{1998AJToumaWisdom}. But the connection is not straightforward. A gas-particle disk would damp to the equator; in this case the disk angular momentum reported in Table~\ref{tab:moon} would be multiplied by the cosine of the inclination to obtain the parallel component, although this is close to 1 in all our cases. For an inclined postimpact state to result in an inclined Moon, rapid coagulation is required, or the presence of nearly lunar-mass postimpact clumps to begin with. Afterward the inclination evolution would depend on the tidal dissipation that drove the Moon's migration \citep[e.g.,][]{2016IcarusChenNimmo}.

\subsubsection{Remnants and debris}

Most material remains gravitationally bound to one of the two largest remnants, for the relatively gentle hit-and-runs considered here, with some exchange across the collisional interface. The total debris mass $m_{\rm esc}$ not bound to either the target or the runner is $\lesssim10^{-3}M_\oplus$ (\papertwo{}), less than a $1/10$ of a lunar mass, in these cases. Although the total energy is lower, more debris is produced in mergers than in slow hit-and-runs. Higher-velocity hit-and-runs and accretions are increasingly dispersive and debris producing, as seen in the right panel of Figure~\ref{fig:phases}.

The structure and size distribution of giant impact debris are uncertain, for reasons similar to why we do not yet know, with any confidence, the structure of the postimpact disk. Like the disk, the escaping debris is an energetic minor fraction of the total colliding mass, whose production is sensitive to geometry, velocity, temperature, EOS, and numerical treatment.
Even if the physics and dynamics of debris production are accurately modeled, resolving $m_{\rm esc}$ explicitly requires quite high numerical resolution. For Earth-mass colliding systems simulated with \num{500000} SPH particles, as in \papertwo{}, clumps smaller than several \SI{100}{\kilo\meter} diameter are not well resolved (a few smoothing lengths). Extensive structures like plumes and spiral arms can be resolved, but not their components.

The total debris mass produced in these hit-and-runs is much smaller than the runner mass, $m_{\rm sr}$, and is therefore ignored for now in the major bodies' dynamical evolution. It would exert some dynamical friction, increasing the tendency toward final accretion. Geologically the production and size distribution of debris are much more significant. One-tenth of a lunar mass is twice the mass of the current main belt, fluxing through the Earth-Moon system on a timescale that overlaps the early solidification and tidal migration timescale of the Moon. It therefore makes a distinguishable difference to early lunar geology whether the returning debris of Moon formation took the form of millions of \SI{10}{\kilo\meter} cratering projectiles, or hundreds of basin-forming bodies, or tens of \SI{1000}{\kilo\meter} bodies, or dust. For example, \citet{2018JGREPerera} showed that if the returning projectiles were large enough to punch through the insulating crust while the Moon was solidifying, the effect would be to greatly accelerate early lunar cooling. Conversely, as noted, one 1000-kilometer impactor could form a remelted hemispheric magma ocean after the Moon had largely solidified.

Returning collisions would most likely happen before the Moon had tidally evolved far away from Earth. Proximity during the bombardment could lead to further mass exchanges and mixing, in addition to the diffusion and equilibration processes described above. Heliocentric impact velocities at the Moon are greater than the escape velocity from Earth at lunar orbit. Earth escape velocity at $R_{\rm Roche}$ is \SI{\sim6}{\kilo\meter\per\second}, almost 3 times the lunar surface escape velocity, so in the case of slow tidal evolution the return bombardment could lead to substantial erosion of the Moon.

Despite its potential significance to the problem of Moon formation, we have not attempted to track the evolution and reimpact of debris in our simulations because it remains a poorly constrained aspect of the problem. Debris production depends sensitively on the details of the collision chain, we are unable to numerically resolve the debris sizes in any case, and including debris adds new dimensions to an already wide-ranging parameter space.

\section{Conclusions}

Moon formation was the final major episode of Earth's accretion, a calamity at the end of the late stage when planetary embryos strayed from their respective feeding zones and collided. Terrestrial planets may have grown pairwise in a series of violent mergers, building up Venus and Earth in perhaps dozens of events, leaving the smaller planets as unaccreted remnants or survivors.

But pairwise accretion does not generally proceed through effective mergers; far from it, hit-and-run collisions are expected to happen about half the time. This means that pairwise accretion proceeds about half the time through hit-and-run return, the process of sequential giant impacts.

\subsection{Pathways to the canonical model}

The goal of this paper is to demonstrate that hit-and-run collisions often lead to terminal mergers that can form a Moon-sized silicate satellite. It builds on prior work.
\paperone{} tracked the dynamical fate of ``runners'' after hit-and-run collisions at 1~AU, focusing on velocities $v_{\rm coll}\lesssim1.2v_{\rm esc}$ that were common in the late stage. A barely escaping runner was shown to recollide with proto-Earth in about \num{\sim0.1}--\SI{1}{\mega\year}, in most cases, with a return velocity close to $v_{\rm esc}$, a terminal merger that is the basis for our hypothesis.

\papertwo{} focused on Earth and Venus, examining the demographics of collision chains. It treated hit-and-run collisions explicitly by applying the surrogate model of \citet{2019ApJCambioni} and \citet{2020ApJEmsenhuberA}. This enabled \textit{N}-body computations to explicitly track the dynamics of targets and runners through sequential collisions. For intermediate-velocity hit-and-runs, with the other major planets included, it was shown that proto-Earth loses a significant fraction of its runners, roughly half, compared to Venus.

\papertwo{} also showed that faster runners, from faster giant impacts into proto-Earth, are most likely to be ultimately accreted by Venus. Also, longer, more complex, more debris-producing chains are relatively common, as are stranded runners. It was seen that proto-Earth could serve as a `vanguard' during late-stage accretion, slowing down interloping major bodies by colliding with them, and delivering their remnants mostly to Venus. Earth, having no such vanguard, would end up comparatively depleted in outer solar system materials.

In an epoch where accretion is occurring overall, $v_{\rm coll}\lesssim1.2v_{\rm esc}$, about half of giant impacts are hit-and-run. Most of those have barely escaping runners that are likely to come back for a terminal merger, hence our proposed scenario where ``proto-Theia'' (mass $\gtrsim m_{\rm imp}\gtrsim0.15 M_\oplus$) collides with proto-Earth at $v_{\rm coll}\sim1.15-1.20 v_{\rm esc}$, where for now we have considered only the most likely impact angle $\theta_{\rm coll}\sim\ang{45}$.
We evaluate the material exchange in this hit-and-run collision and obtain the target's final spin rate and the runner's mass and egress velocity. The target and runner evolve dynamically until there is a second collision, which can resemble the canonical scenario.

A faster initial hit-and-run with proto-Earth is possible, but as noted, these runners are more likely to terminate at Venus (\papertwo{}). A hit-and-run return starting with a faster projectile, for example from the outer solar system, is therefore a less likely scenario for Moon formation around the Earth than it would be for Venus. While our scenario allows for higher (and, we argue, more realistic) initial relative velocities than the canonical model, it favors a barely escaping runner and hence a proto-Theia deriving from the inner solar system.

Just as $\theta_{\rm imp}\ang{\sim45}$ is the most probable impact angle, $\varphi\ang{\sim90}$ is the most probable alignment angle. Orthogonal return collisions are the norm. We find that these lead to disk masses and compositions that are consistent with Moon formation. Moreover, two collisions in a row lead to improved isotopic equilibration, by about \SI{10}{\percent} in simulations, and a substantially inclined postimpact disk.

\subsection{Further Reaccretion and Mixing}

The return timescale of runners and debris, of order 0.1--1~Myr, is shorter than most error bars on geochronometry applied to the Moon-formation era. A hit-and-run return and its consequences would therefore be resolvable in relative time but not absolutely. It is therefore important to understand the sequential events of follow-on collisions and material exchanges, which altogether constitute the giant impact.

Our nominal scenario improves Earth-Moon isotopic similarity but is not a sufficient explanation.
It is not yet clear whether proto-Earth and proto-Theia must be so perfectly equilibrated \citep[e.g.,][]{2020NatGeoCano}, and $\delta f_\mathrm{T}^\mathrm{c}=0$ is too high a bar to attain. For now we note that our scenario is an improvement and also that our methodology underestimates the degree of compositional equilibration for several reasons.

Numerical effects in SPH could limit the particle exchanges during collisional shearing in the simulated hit-and-runs.
These could likewise limit proto-Earth and Theia exchanges in simulations of the canonical model, so that accounting for more mixing would improve canonical models as well. If mixing is suppressed generally by SPH in giant impacts, it would accumulate more significantly in our two-collision model.

Also, we have ignored that the runner is stripped of a fraction of its mantle by the hit-and-run. Theia ends up \num{\sim10}--\SI{20}{\percent} more metal-rich than proto-Theia in our starting cases (Table~\ref{tab:hrc}), yet with no knowledge of proto-Theia we use a chondritic runner in every case to represent the terminal collision.
Because the two major bodies remain on close orbits, the target ends up accreting the majority of the runner's stripped silicates, according to its greater mass \citep{2014NatGeoAsphaug}. Consequently, Theia's overall contribution to the Moon is less than we compute.

Another important consideration is that the postimpact middle-Earth might retain a magma ocean from the hit-and-run, given that the interlude timescale may be much shorter than the cooling timescale. From SPH simulations of proto-Earth targets with deep magma oceans, \citet{2019NatGeoHosono} obtain more massive protolunar disks with significantly greater proto-Earth mass fraction. Slow hit-and-runs are not as effective as mergers at producing magma oceans, however \citep{2021EPSLNakajima}.

Additionally, we neglect any blending of material between the newly formed Earth and Moon, caused by the reimpact of escaping materials ranging in size from dust to diverse large bodies \citep[e.g.,][]{2017EPSLGenda}. An era of strong dynamical exchanges will result \citep[e.g.,][]{1993IcarusGladman,2012MNRASJacksonWyatt}, and a fraction of returning debris would strike the Moon, potentially to strip away, overturn, or remelt its outer layers. Most would strike the Earth, potentially to layer the lunar crust with its mega-ejecta. Such effects apply to the canonical model, and more so in collision-chain scenarios that produce more debris.

More violent hit-and-runs than we have studied, at higher velocity and smaller impact angle \citep{2012IcarusReufer}, produce much more debris and would increase the significance of these effects. Catastrophic disruption of the projectile, at velocities faster than about $1.3v_{\rm esc}$ (leaving the target mostly intact), would produce a deeply mantle-stripped metallic runner, or multiple runners (right panel of Figure~\ref{fig:phases}), with half or more of the projectile escaping as variegated remnants and debris. Longer, more energetic collision chains would lead to more thorough mixing, although at some point perhaps not a Moon. A higher-velocity proto-Theia would originate farther from proto-Earth, counteracting the desired tendency toward isotopic similarity. Also, as noted, according to \papertwo{} Earth is more likely to lose its faster runners to Venus than to retain them, in which case Moon formation would not occur at Earth.

Longer chains, multiple runners, and masses of lesser remnants leave us with many scenarios to consider, as well as the dynamics, chronology, and geochemistry those entail---scenarios for protolunar disk formation and, afterward, for the survival and evolution of the disk and early Moon in the face of returning remnants. And each scenario provides physical and geochemical consequences that might be confirmed or refuted by the earliest lunar geology.

Here we have only analyzed the slow and simple cases without much debris. Considering the other extreme, the most obvious upper limit on debris production is that the Moon exists. Debris-laden chains may be dominated by a Mars-mass Theia whose return collision produces the Moon, but if there are several lunar masses of accompanying debris, their return bombardment over the next $\sim1$~Ma might destabilize or destroy the new-formed Moon.

\subsection{Why Earth?}

Our hypothesis for Moon formation begins with two terrestrial planets colliding at $\lesssim1.2v_{\rm esc}$, which for expected impact angles and mass ratios is a hit-and-run collision. A return collision often happens some \num{0.1}--\SI{1}{\mega\year} later, much slower, and unaligned to the first. Proto-Earth finally acquires the runner and its angular momentum, producing a lunar-mass disk of mixed silicate composition, in the common case of an impact angle \ang{\sim45}. This provides a dynamical pathway to the canonical scenario and has the additional advantage of causing greater isotopic mixing.

But it is not without conundrums of its own. It was found in \papertwo{} that proto-Venus is more effective than proto-Earth at ultimately accreting the runners of hit-and-run collisions. In fact, Venus is the terminus for more than half of proto-Earth's faster runners. Earth mainly holds on to its slower runners, which motivates our present focus on low-velocity hit-and-run events. For faster hit-and-runs, a disk-forming terminal graze-and-merge collision would have been more probable at Venus. This shifts the question from \textit{Why does the Earth have a moon?} to \textit{Why doesn't Venus have a moon?}

One approach is to invoke the small number statistics of giant impacts. Maybe Venus did not suffer any giant impacts \citep{2017EPSLJacobson}. Or maybe the final giant impact into Venus was head-on, wiping out any satellite from before, instead of the requisite graze-and-merge collision. If the terrestrial planets formed directly by pebble accretion, maybe there was only one giant impact, Theia and proto-Earth \citep{2021SciAdvJohansen}. But it would have to be slow, nearly $v_{\rm esc}$, to form the Moon by the demonstrated mechanism of a graze-and-merge collision. A faster collision, especially between bodies of comparable mass, would likely be a hit-and-run collision. While this could set the stage for a Moon-forming return collision such as we propose, a fast runner from the Earth is more likely to end up at Venus. So this is not an immediate solution to our conundrum.

While Venus may have been more likely than the Earth to have acquired a massive satellite by our hypothesis, it may also have been more likely to have lost one. For the same reason that Venus reaccretes a greater fraction of its runners, compared to Earth, it also reaccretes a greater fraction of its giant impact debris, of which it produces more for a given projectile (the mass ratio being larger, and the denominator $v_{\rm esc}$ being smaller in Figure~\ref{fig:phases}). The orbit being smaller, 0.7~AU, the debris will recollide with Venus sooner, with greater efficiency, and at higher velocity. So the argument, that returning debris would erode or even destroy a massive satellite after its formation, is more relevant to Venus than Earth.

The interconnectedness of terrestrial planet formation demonstrated in this set of papers points to the significance of understanding the solidification, layering, and early excavation geology of the Moon, with its profound record of the aftermath of its formation, and of obtaining surface samples from Venus, which, if there was a late stage, should have a commonality with Earth.

\begin{acknowledgments}
E.A., A.E., S.C., and S.R.S. were supported by NASA grant 80NSSC19K0817, ``Application of Machine Learning to Giant Impact Studies of Planet Formation.''
Allocation of computer time from the University of Arizona center of High Performance Computing (HPC) is gratefully acknowledged.
The authors thank Kevin Zanhle for his insights, as well as two anonymous referees, whose detailed efforts helped us clarify our research and improve its context.
We are especially indebted to Jay Melosh (1947-2020) for his insightful teachings about planetary geology and formation and his singular enthusiasm for giant impacts and the physics of collisions.
\end{acknowledgments}

\software{matplotlib \citep{2007CSEHunter}, \texttt{collresolve} \citep{2019SoftwareEmsenhuberCambioni}}

\bibliographystyle{aasjournal}
\bibliography{manu,add}

\end{document}